\documentclass[pra,twocolumn, floatfix, footinbib,superscriptaddress]{revtex4-1}
\usepackage[utf8x]{inputenc}
\usepackage[english]{babel} 
\usepackage[english]{babel} 
\usepackage[colorlinks=true,bookmarks=false,citecolor=blue,urlcolor=blue]{hyperref}

\usepackage{amsmath,amssymb,esint}
\usepackage{graphicx}
\usepackage{verbatim}
\usepackage{color}
\usepackage[english]{babel}
\usepackage{bm}
\usepackage{natbib}
\usepackage[normalem]{ulem}
\usepackage{epstopdf}
\usepackage{multirow}

\usepackage{ragged2e}

\def\dfrac{\displaystyle\frac}
\renewcommand {\phi}{\varphi}

\newcommand{\e}{{\rm e}}
\newcommand{\rmd}{{\rm d}}
\newcommand{\rmi}{{\rm i}}
\newcommand{\eps}{\varepsilon}

\newcommand{\rot}{\mathop{\mathrm{rot}}\nolimits}

\usepackage{amsmath}
\usepackage{graphicx}
\usepackage[colorinlistoftodos]{todonotes}

\newcommand{\om}{\omega}
\newcommand{\bee}{\begin{equation}}
\newcommand{\ene}{\end{equation}}

\newcommand{\bea}{\begin{eqnarray}}
\newcommand{\ena}{\end{eqnarray}}

\newcommand{\bt}{BaTiO$_3$}
\newcommand{\ga}{AlGaAs}

\newcommand{\red}[1]{{\color{black}#1}}
\newcommand{\blue}[1]{{\color{black}#1}}
\newcommand{\bluee}[1]{{\color{black}#1}}

\newcommand{\ve}[1]{{\bf {#1}}}
\newcommand{\te}[1]{{\widehat{#1}}}
\graphicspath{{pics/}}

\let\vec=\mathbf

\begin{document}
\title{Second-harmonic generation in Mie-resonant dielectric nanoparticles made of noncentrosymmetric materials}

\author{Kristina Frizyuk}
\affiliation{ITMO University, St. Petersburg 197101, Russia}

\author{Irina Volkovskaya}
\affiliation{Institute of Applied Physics, Russian Academy of Sciences, Nizhny Novgorod 603950, Russia}

\author{Daria Smirnova}
\affiliation{Institute of Applied Physics, Russian Academy of Sciences, Nizhny Novgorod 603950, Russia}

\author{Alexander Poddubny}
\affiliation{ITMO University, St. Petersburg 197101, Russia}
\affiliation{Ioffe Institute, St.~Petersburg 194021, Russia}

\author{Mihail Petrov}
\affiliation{ITMO University, St. Petersburg 197101, Russia}

\begin{abstract}
We develop a multipolar theory of second-harmonic generation (SHG) by dielectric nanoparticles made of noncentrosymmetric materials with bulk quadratic nonlinearity. We specifically analyze two 
regimes of optical excitation: illumination by a plane wave and single-mode excitation, 
when the laser pump drives the magnetic dipole mode only. 
Considering two classes of nonlinear crystalline solids (dielectric perovskite material and III-V semiconductor), we apply a symmetry approach to derive selection rules for the multipolar composition of the nonlinear radiation.  
The developed description can be used for design of efficient nonlinear optical nanoantennas with reconfigurable radiation characteristics.  
\end{abstract}



\maketitle

\section{Introduction}

The resonant response is one of the main routes to increase the efficiency of nonlinear signal generation at the subwavelength scales in the absence of phase matching effects. That is why optical nonlinearity at the nanoscale is usually associated with the enhancement of electric fields in plasmonic nanostructures due to geometric plasmon resonances~\cite{Kauranen2012,Panoiu2018}. Despite  the significant progress in this area \cite{Butet2015}, there exist fundamental drawbacks that limit the efficiency of nonlinear generation with metallic structures. Besides the evident problem of high ohmic losses, typical metals have cubic lattice with inversion  symmetry which restricts second-order nonlinear effects, such as the second-harmonic generation (SHG)~\cite{Boyd2003}. It can be observed only due to  the surface effects or the field gradients in the bulk of nanoparticles~\cite{Butet2015, Capretti2014}, which are relatively weak. Recently, a novel nanophotonic platform  based on high-index dielectric nanoparticles has emerged~\cite{kuznetsov2016science}. All-dielectric nanostructures  are free from high ohmic losses, and offer wide variety of  dielectric and semiconductor materials including those with nonzero bulk second order susceptibility  tensor.  Excitation of Mie resonances in such nanoparticles provides novel opportunities for  nonlinear optics~\cite{Smirnova2016,Kruk2017_ACS_Phot}, and allows one to achieve record-high nonlinear conversion efficiencies at the nanoscale~\cite{Camacho-Morales2016,Cambiasso2017,Vabishchevich2018,Gili2018,Rocco2018,Carletti2015,Poddubny2018arXiv}.

Despite the intense experimental stuides of the SHG effects in Mie-resonant nanostructures, a comprehensive theory of the SHG emission from  nanoparticles with nonzero bulk nonlinearity tensor $\hat\chi^{(2)}$ has not been proposed yet. The important works related to the SHG generation were focused on the surface and bulk effects in  nanoparticles with centrosymmetric crystalline lattice: in noble metal  nanoparticles \cite{Dadap, Pavlyukh2004, Mochan} including the shape effects \cite{Finazzi2007}, and in Mie-resonant silicon nanoparticles \cite{Makarov2017,Smirnova2018}. In this work, we theoretically study the SHG by individual spherical high-index dielectric nanoparticles made of non-centrosymmetric materials  (aluminum gallium arsenide AlGaAs and barium titanate BaTiO$_3$), which possess a large bulk quadratic susceptibility. These materials are actively employed for nonlinear all-dielectric nanophotonics~\cite{Camacho-Morales2016, Kruk2017, Timpu2017, Ma2017,Vabishchevich2018}. 
We systematically describe the SHG in nanoparticles and mechanisms of its resonant enhancement, depending on the symmetry of the crystalline structure and polarization of the incident light.
We employ methods of multipolar electrodynamics providing a transparent interpretation for the measurable far-field characteristics, such as radiation efficiency and radiation patterns~\cite{Smirnova2016,Smirnova2016ACSPh,Smirnova2018}. 

Using analytical techniques, we demonstrate the ability to manipulate the nonlinear radiation  of a spherical nanoparticle by varying illumination properties. By means of symmetry analysis of the SHG process we obtain the selection rules for the nonlinear generation, and identifiy which channels of multipole  composition are active in SHG. These rules previously were known only for nanoparticles of a spherical~\cite{Dadap} and arbitrary shape~\cite{Finazzi2007} made of centrosymmetric materials. The knowledge of  these basic  mechanisms of nonlinear generation in a single spherical nanoparticle can be extended in application to complex nonlinear structures, such as nanoparticle oligomers \cite{Martin2013,Timpu2017a} or nanoparticle arrays in metasurfaces \cite{Kruk2017a}.

The paper is organized as follows: in Sec.~\ref{sec:II-plane+single} we discuss the problem of nonlinear light scattering of a plane wave by a dielectric nanoparticle made of \bt\ {or} \ga \ materials. Applying  Green's function approach, we calculate the efficiency of SHG and multipolar content of the second-harmonic (SH) field. We also consider the particular case of SHG through excitation of a single magnetic dipole mode. We discuss how the intensity and the far-field properties of the SH field vary while direction of the excited dipole moment changes relatively to the crystalline structure of material. By explicit calculations we show how the mode content of SH field varies. In Sec.~\ref{Sec:Symmetry} we derive selection rules which govern the channels of mode coupling at fundamental and SH wavelengths based on the symmetry of vector spherical harmonics and the crystalline structure. In Sec.~\ref{Sec:Discussion} we apply the formulated selection rules to explain the results obtained in Sec.~\ref{sec:II-plane+single} and build the complete table of possible generated multipoles for SHG process driven by dipole modes. 

\section{Second-harmonic generation formalism} 
\label{sec:II-plane+single}
\subsection{Green's function approach. Plane-wave excitation} 
\label{sec:II-plane}
We consider a  spherical dielectric particle of the radius~$a$
characterized by a frequency-dependent dielectric permittivity $ \varepsilon_2(\omega) $, embedded in a homogeneous host medium with $\varepsilon_1=1$. The nanoparticle is made of the material with a noncentrosymmetric crystalline structure, and its nonlinear electromagnetic properties are captured by the quadratic susceptibility tensor $\hat \chi^{(2)}$. \bluee{
While for AlGaAs the linear susceptibility tensor $\varepsilon_2(\omega)$ is isotropic, for  BaTiO$_{3}$ this tensor inherits the  uniaxial crystal structure of the material. In this case,  the anisotropy of SHG tensor and of the linear permittivity tensor are not two independent phenomena and have a common microscopic origin~\cite{Pavlyukh2012}. However the anisotropy is rather weak in the case of BaTiO$_3$, it dramatically increases the complexity of the problem comparing to  isotropic linear scattering. Hence, from now on we use the approximation of isotropic linear susceptibility tensor. The effect of anisotropy on the selection rules will be  discussed in more details at the end of the paper in Sec.~\ref{Sec:Discussion}.}


The problem of linear light scattering by a sphere is solved using the multipolar expansion following the Mie theory~\cite{Bohren, Mie}. 
In our work we consider time dependence of the fields  in the form $\e^{-i\omega t}$. For the  illumination by the $x$-polarized plane wave $E_{0}\bm e_{x}\e^{\rmi k_{1}z}$ incident along the $z$ direction, the field inside the spherical nanoparticle ($r<a$) is expanded in vector spherical harmonics as follows:

\begin{multline}\label{planewave}
\vec E^{\om}(\vec r) =\sum_{n=1}^\infty i^n\frac{(2n+1)}{n(n+1)}\left[ c_n \vec{M}^{(1)}_{o1n}(k_2 (\omega),\vec r) \right.\\ - i \left.d_n\vec{N}^{(1)}_{e1n}(k_2 (\omega),\vec r)\right]\:,
\end{multline}
where the wavenumbers $k_1(\omega)={\omega}\sqrt{\varepsilon_1}/c$, $k_2(\omega)={\omega}\sqrt{\varepsilon_2(\omega)}/c$. Magnetic  $\vec{M}_{o1n}$ and electric $\vec{N}_{e1n}$ spherical harmonics  with the total angular momentum $n$ and the momentum projection $\pm 1$, the indexes $e,o$ describing their parity with respect to the reflection along $y$ axis (or $\varphi \rightarrow -\varphi$ transformation), and the coefficients $c_n$, $d_n$ are given in  Appendix~\ref{app:definitions}, the superscript $(1)$ is used to  define spherical Bessel functions.

 \begin{figure}[t!] 
	\includegraphics[width=0.8\linewidth]{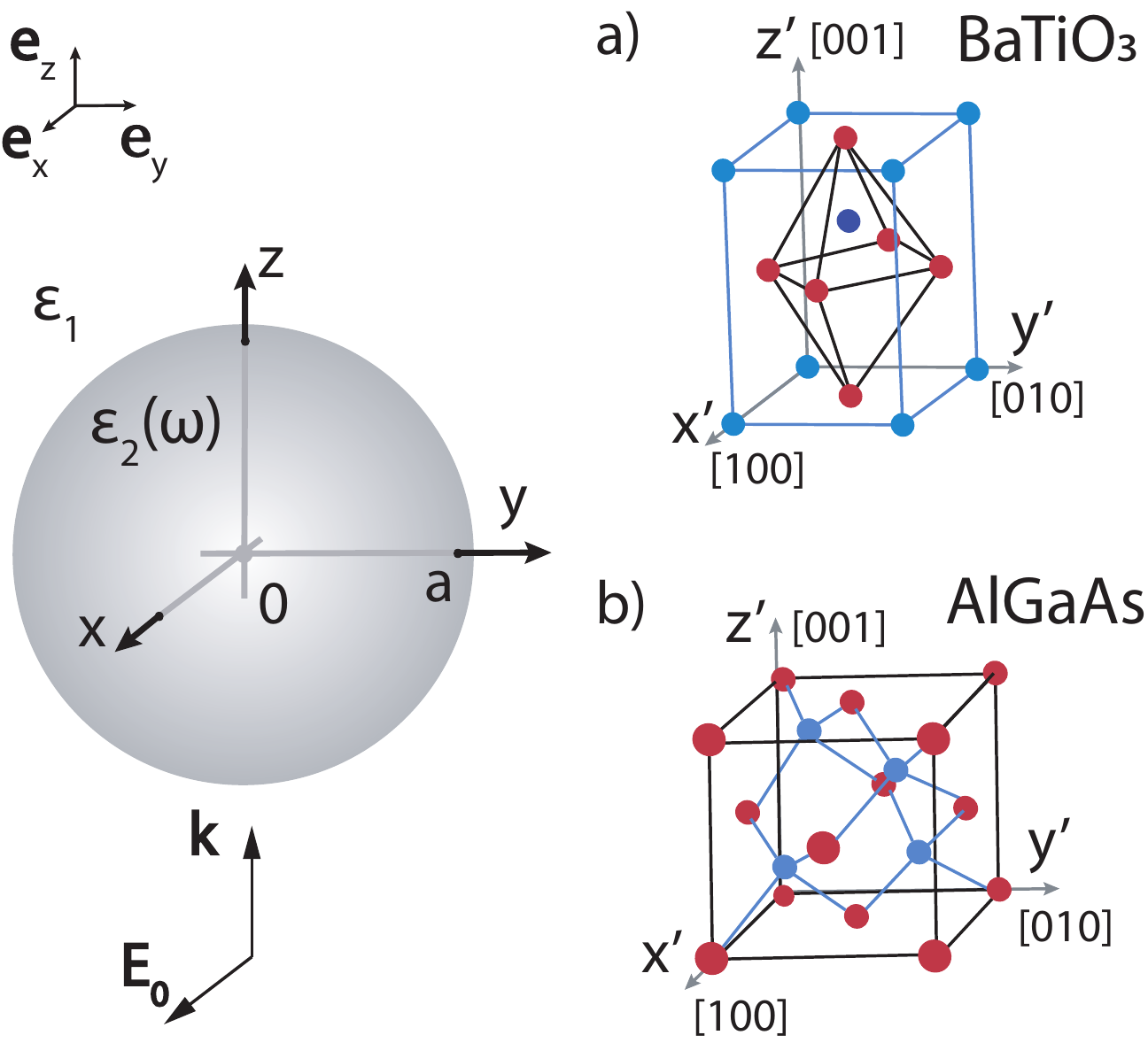} 
	 \caption{Geometry of the considered problem. The  crystalline lattice structure of materials under the consideration, \bt (a) and AlGaAs (b), is also shown in figure. The orientation of crystalline lattice is with respect to the coordinate system is fixed throughout the paper if else is not specified.  } \label{figBTgeome} 
\end{figure}

The induced nonlinear polarization at the second-harmonic frequency is defined by the second-order polarizability tensor:
\begin{equation} \label{poln}
P_\alpha^{2\om}{(\vec r)}= \varepsilon_0 \chi^{(2)}_{\alpha\beta\gamma} E^{\om}_\beta( \vec r) E^{\om}_\gamma(\vec r) ,
\end{equation}
where $E^{\om}_{\alpha}$ are the Cartesian components of the fundamental field inside the nanoparticle. We take into account only the bulk nonlinear response leaving outside the consideration  potential surface sources of SHG. The SH field 
outside the particle at $r>a$ can be found using  dyadic Green's function $\te {{\bf G}}$ of a sphere: 

\begin{equation}\label{eq:Eout}
\vec{\bf E}^{2\om}({\bf r})=(2\omega)^2\mu_0\int\limits_V dV' \te {{\bf G}}(2\omega,{\bf r,r'}) \ve P^{2\omega} ( \ve r')\:,
\end{equation}
satisfying  the following equation 
$\rot\rot {\bf \hat G}(\omega,\vec r,\vec r')=\left(\frac{\omega}{c}\right)^{2}\eps(\ve r, \om){\bf \hat G}(\omega,\vec r,\vec r')+ {\bf \hat 1}\delta(\bm r-\bm r')\:,
$
where $\hat{\bf 1}$ is the unit dyadic, $\eps(\ve r, \om)=\eps_{2} (\om)$ for $r<a$, and $\eps(\ve r)=1$ for $r>a$.

The explicit multipole decomposition of the Green function is given in Ref.~\onlinecite{Xiang1996}, and also in Appendix ~\ref{app:definitions}. Substituting the expansion of the Green function in the form of the spherical waves into Eqs.~\eqref{poln},\eqref{eq:Eout} we obtain the multipolar decomposition of the second harmonic field 


%

\begin{multline}
 \vec E^{2\omega}(\vec r) =
 \sum_{n=1}^\infty \sum_{m=0}^n\sum_{W=M,N}
 E_0( D_{Wemn} \vec{W}_{emn}^{(3)}[k_1(2\omega),\vec r]+\\
 + D_{Womn} \vec{W}_{omn}^{(3)}[k_1(2\omega),\vec r])\:.\label{eq:E2omega}
\end{multline}
Here, the denotation $W=M,N$ distinguishes between electric and magnetic harmonics, the superscript $(3)$ is used to define spherical Hankel functions of the first kind. The expansion coefficients
$D_{W^e_omn} $
are readily evaluated as a sum of overlap integrals between the two vector spherical harmonics at the $\omega$ frequency and 
one harmonic at the  $2\omega$ frequency,  weighted by the $\te \chi^{(2)}$ tensor:
\begin{equation}
D_{W_{^e_o mn}}\sim \int\limits_V  {\vec{W}}^{(1)}_{^e_o mn}[k_2(2\omega),\vec r] \te\chi^{(2)} \vec E^{\om}(\vec r) \vec E^{\om}( \vec r) dV\:.\label{eq:Dcoeff}
 \end{equation}
The explicit form of these coefficients is given in Appendix \ref{app:definitions}. 

Finally, the second-harmonic conversion efficiency $\sigma_{\text{SH}} $, defined as the ratio of the total SH radiated power $\mathcal{P}_{SH}$ to the energy flux of the fundamental wave $I_0$ through the geometrical cross section $\pi a^2$ of the particle, can be expressed through the coefficients $D_{W^e_o mn} $ as follows \cite{Fuller1994}:
 \begin{multline} \label{total}
\sigma_{\text{SH}} =\dfrac{\mathcal{P}_{SH}}{\pi a^2I_0}=\frac{2\pi}{\pi a^2 [k_1(2\omega)]^2}
\sum_{n=1}^\infty\sum_{W=M,N}
\frac{n(n+1)}{(2n+1)}\times\\\times\Bigl[
\sum\limits_{m=1}^{\blue n}\frac{(n+m)!}{(n-m)!}(|D_{Wemn}|^2+|D_{Womn}|^2)+
2|D_{We0n}|^2
\Bigr].
\end{multline}

Using Eq.~\eqref{total}, we calculate the conversion efficiency $\sigma_{\text{SH}}$ 
for BaTiO$_3$ and AlGaAs particles of subwavelength sizes under the plane-wave excitation. In this section, we assume that the main axes of crystalline lattice are oriented along the coordinate system: $[100] \|  \ve{e}_x,\ [010] \| \ve{e}_y, \ [001] \| \ve{e}_z$ (see Fig.~\ref{figBTgeome}).  In Section~\ref{Sec:Discussion} we will discuss other crystalline orientations. 
In the chosen coordinate system the second-order polarization of the BaTiO$_3$ crystal  has the following form:
\begin{equation}
\left( \begin{array} {ccc}
P_x^{2\omega}\\
P_y^{2\omega}\\
P_z^{2\omega} \end{array} \right)\!=\!
\varepsilon_0 \!\left( \begin{array} {cccccc}
0 & 0 & 0 & 0 & d_{15} & 0  \\
0 & 0 & 0 & d_{15} & 0 & 0 \\ 
d_{31} & d_{31} & d_{33} & 0 & 0 & 0\end{array} \right)
\!
\left( \begin{array} {cccccc}
E^{\omega}_xE^{\omega}_x\\
E^{\omega}_yE^{\omega}_y \\ 
E^{\omega}_zE^{\omega}_z \\
2E^{\omega}_y E^{\omega}_z  \\
2E^{\omega}_xE^{\omega}_z  \\
2E^{\omega}_xE^{\omega}_y \end{array} \right)\label{eq:chiBT},
\end{equation}
where $\chi_{zzz}=d_{33}=6.8$~pm/V, $\chi_{zxx}=\chi_{zyy}=d_{31}=15.7$~pm/V, $\chi_{
xxz}=\chi_{yyz}=d_{15}=17$~pm/V~\cite{Cabuk2012}. In the principal axis system of the AlGaAs crystal, the tensor of the second-order nonlinear susceptibility contains only off-diagonal  elements $\chi^{(2)}_{ijk}\equiv \chi^{(2)}=100$~pm/V being non-zero if any of two indices $i, j, k$  do not coincide: 
{
\begin{equation}\label{eq:chiAlGaAs}
\left( \begin{array} {ccc}
P_x^{2\omega}\\
P_y^{2\omega}\\
P_z^{2\omega} \end{array} \right)\!=\!
2 \varepsilon_0 \chi^{(2)} 
\left( \begin{array} {ccc}
E^{\omega}_y E^{\omega}_z  \\
E^{\omega}_x E^{\omega}_z \\
E^{\omega}_xE^{\omega}_y  \end{array} \right)\:.
\end{equation}}

  The fundamental wavelengths are fixed for BaTiO$_3$ and AlGaAs  to 1050 nm and 1550 nm respectively. These values were chosen in accordance with the typical experimental frequencies used for observation of SHG from these materials and  correspond to the Yb$^{+3}$ laser (1050 nm) \cite{Timpu2017} or the Er$^{+3}$ doped fiber laser (1550~nm)\cite{Camacho-Morales2016}. Since AlGaAs has a higher refractive index ($\sim 3.5$) compared to BaTiO$_3$ ($\sim 2.4$), the particle sizes are within the same range.   

	\begin{figure}[!t]\begin{center}
	 \includegraphics[width=0.95\linewidth]{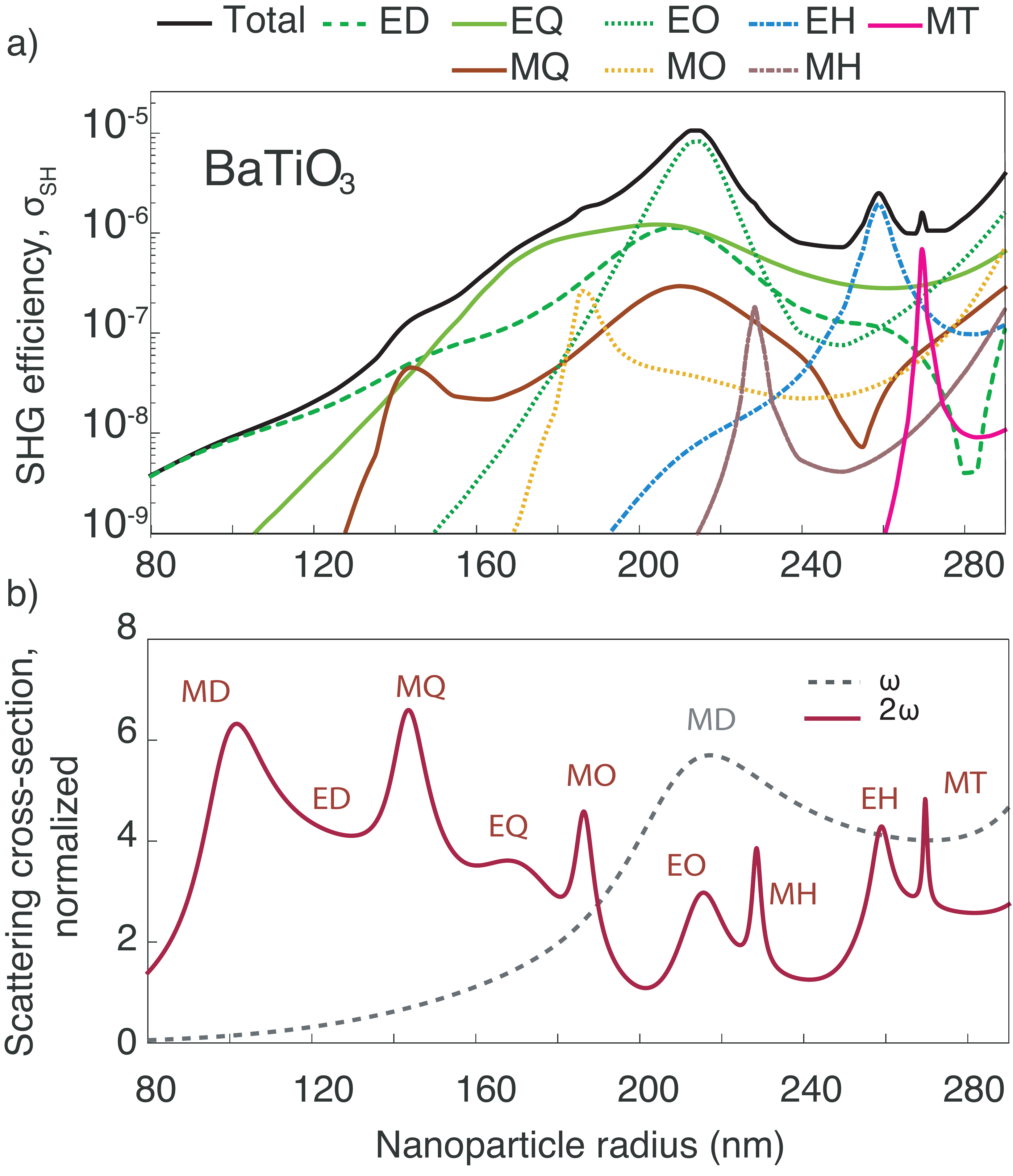}
	\caption{Second harmonic generation spectra spectra for BaTiO$_{3}$ calculated at the normal incidence, according to Fig.~\ref{figBTgeome}. Incident wave wavelength is 1050 nm. (a) Solid black line shows the total SH intensity, normalized to the incident power  $I_0=10^{13}$~W/m$^2$ and the geometric cross-section $\pi a^2$. Colored lines show different multipole contributions to the second harmonic field. (b) Scattering cross section, normalized to the geometric cross-section for the two wavelengths: 1050 nm (dashed line) and 525 nm (solid line), to show the positions of the multipole resonances. \blue{ E/MD - electric/magnetic dipole, Q- quadrupole, O - octupole, H - hexadecapole, T - triacontadipole.}} 
	\label{btmult}\end{center}
\end{figure}

\begin{figure}[!t]
	\includegraphics[width=0.95\linewidth]{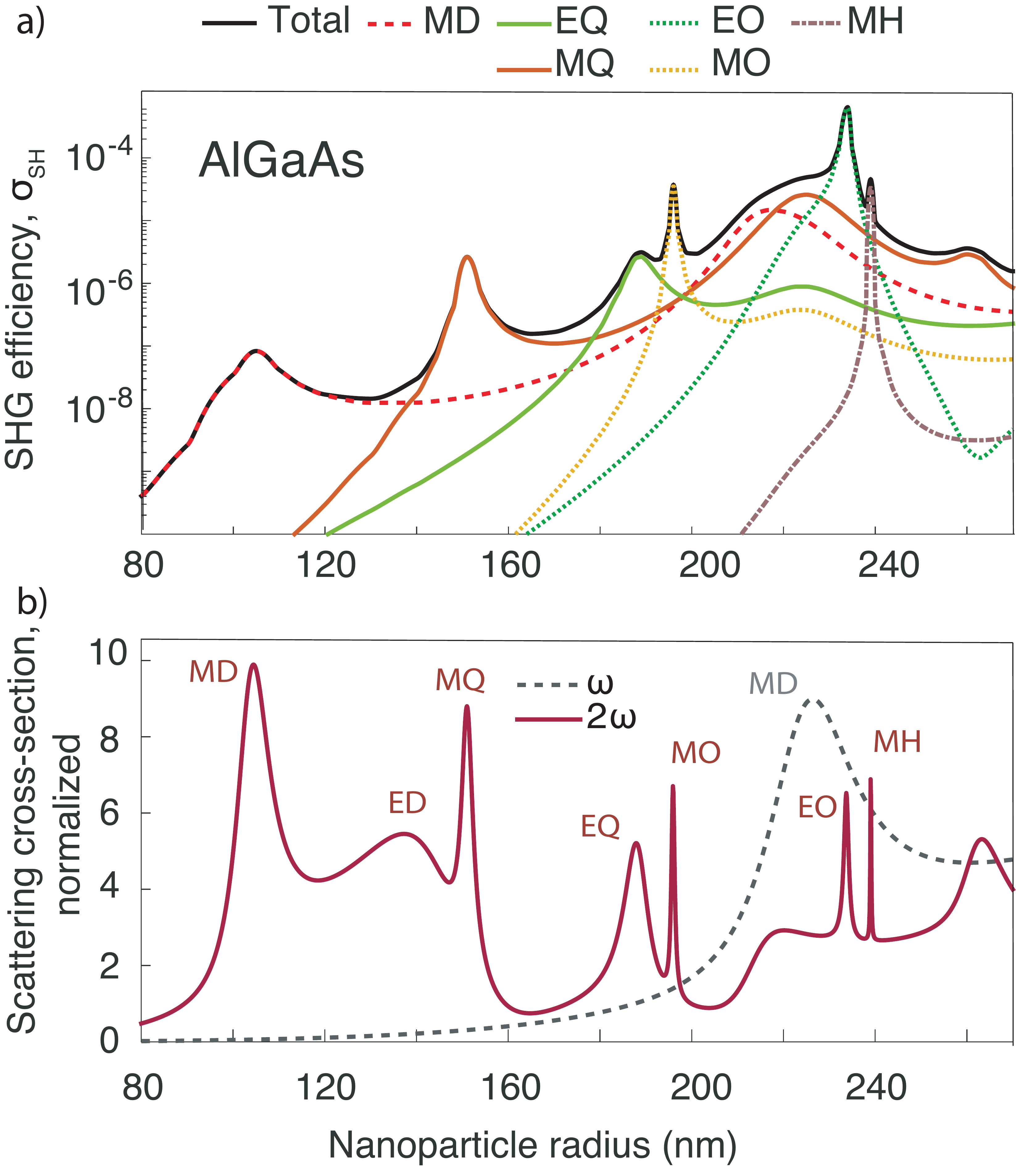}
	\caption{Second harmonic generation spectra for AlGaAs at the normal incidence, according to Fig.~\ref{figBTgeome}. Incident wavelength is 1550 nm. (a) Solid black line shows the total SH intensity, normalized to the incident power  $I_0=10^{13}$~W/m$^2$ and the geometric cross section $\pi a^2$. Colored lines are multipoles contributions into the SH intensity. (b) Scattering cross section, normalized to the geometric cross section for the two wavelengths: 1550~nm (dashed line) and 775~nm (solid line), to show the positions of multipole resonances. \blue{E/MD - electric/magnetic dipole, Q- quadrupole, O - octupole, H - hexadecapole, T - triacontadipole.}} 
	\label{dalgamul}
\end{figure}
 
The calculated dependences of SHG on the nanoparticle radius are shown in Figs.~\ref{btmult}(a), and \ref{dalgamul}(a) demonstrating pronounced resonant structure. In order to distinguish between different multipolar resonances,
we have separately calculated the contribution from each multipolar channel in Eq.~\eqref{total} [see the colored curves in Fig.~\ref{btmult}(a), and Fig.~\ref{dalgamul}(a)]. The contributions of the harmonics with the same total angular momentum $n$ and different momentum projections $m$ are combined together. Identification of the harmonics with particular momentum projection contributing to the SH emission will be discussed in detail in Sec.~\ref{Sec:Discussion} (see  Tables~\ref{BTztable}, \ref{BTxtable}, \ref{AlGatable}). We also support the SH spectra with the  plot of  the linear  scattering spectra of a plane wave  at the fundamental and SH wavelenghts in Figs.~\ref{btmult},\ref{dalgamul}~b), clearly showing individual Mie resonances.

One can see that the peaks at the SH wavelength are modulated with the broad resonance at the fundamental wavelength, which is clearly seen by comparing the panels (a)  and (b)  in Fig.~\ref{btmult} and Fig.~\ref{dalgamul}. As a result, the SHG efficiency increases by several orders of magnitude when approaching the magnetic dipole (MD) resonance at the fundamental wavelength \cite{Chervinskii2018,Scherbak2018}. 
The dramatic enhancement is observed when the double-resonance condition is fulfilled  ~\cite{Celebrano2015,Thyagarajan2012,Yang2017a} for instance at electrical octupole (EO) resonance [see Figs.~\ref{btmult}, and \ref{dalgamul} (a)]. Results of our analytical calculations are confirmed by full-wave numerical modeling performed with the finite-element solver COMSOL Multiphysics, following the procedure applied in Refs.~\cite{Smirnova2016ACSPh,Camacho-Morales2016,Kruk2017,Smirnova2018}. The multipolar amplitude coefficients are then numerically retrieved 
and reproduce Figs.~\ref{btmult},\ref{dalgamul}.

The magnitude of the SH conversion efficiency is intensity-dependent $\sigma_{\text{SH}}\sim I_0$, as it describes the two-photon process. Specifically, for a given intensity of the incident wave of $I_0=1$ GW/cm$^2$ the conversion efficiency reaches the value of $10^{-5}$ for \bt \ nanoparticle and $5\cdot 10^{-4}$ for \ga \ nanoparticles in the same radius range of around 200~nm. These values are  about one order of magnitude higher than the experimental values  measured for nanodisks in similar experimental conditions~\cite{Gili2018,Camacho-Morales2016}. This discrepancy can be related to the lower SHG efficiency from disk resonators studied in the experiments due to the substrate effects and the uncertainty of retrieving of the efficiency value from experimental data.

Another important feature is the particular multipolar content of the SH field. For instance one can notice that the MD  is absent in the SH field generated in the \bt \ nanoparticle, and no electric dipole (ED) field is generated in the \ga \  nanoparticle. This cancellation is dictated by the symmetry of the $\te \chi^{(2)}$ tensor and direction and polarization of the fundamental wave. It   will be further  illustrated in Sec.~\ref{Sec:Single_mode}, studied in detail from the symmetry point of view in Sec.~\ref{Sec:Symmetry}, and discussed in Sec.~\ref{Sec:Discussion}.  
%
%

\subsection{Single-mode approximation} 
\label{Sec:Single_mode}
Here, we specifically focus on the SHG driven by the MD mode only.
In the vicinity of pronounced resonances, the field distribution inside the particle excited by the fundamental wave can be approximated by the corresponding eigenmode~\cite{Smirnova2016ACSPh,Smirnova2018}. Selective and enhanced coupling to specific multipole modes can be facilitated by the beam engineering~\cite{Das2015,Melik-gaykazyan2018}.\blue{ If the refractive index is high enough, $k_1(\omega) a\sqrt {\varepsilon_2(\omega)}  \sim \pi$ , the fundamental MD resonance dominates in the fundamental field in particular spectral region (around 220~nm radius for the fundamental wavelengths in Fig.~\ref{btmult}, and Fig.~\ref{dalgamul}).} The case of SHG driven by MD excitation represents an instructive example for understanding the multipolar nature of the generated electromagnetic fields in Mie-resonant dielectric nanoparticles.

\begin{figure}[!t] 
\includegraphics[width=0.8\linewidth]{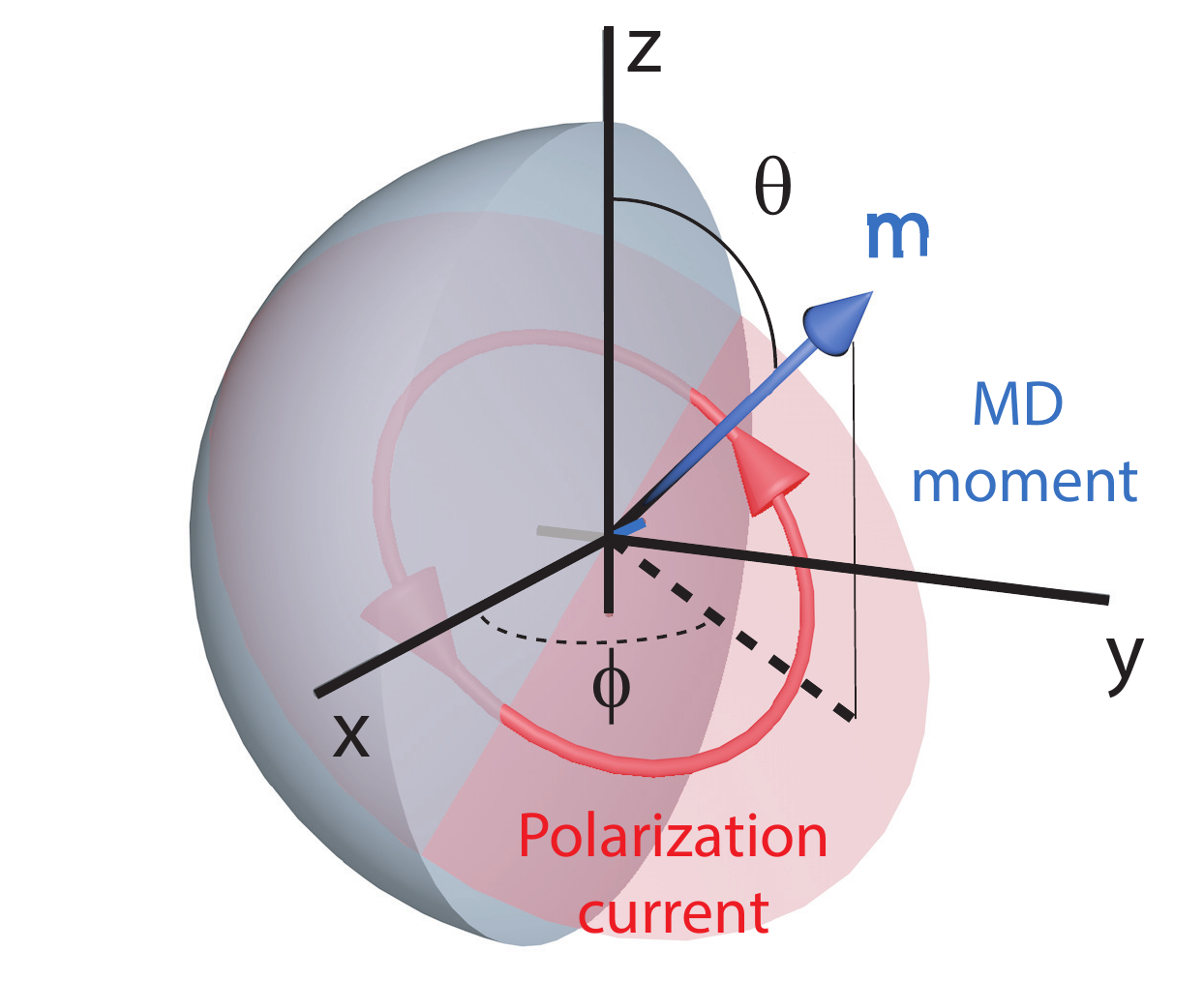}
\caption{The geometry of the  magnetic dipole mode excitation in the nanoparticle and corresponding angles of dipole moment rotation. } \label{fig:MDpump}
\end{figure}
We employ a single-mode approximation and assume that the field inside the \ga \  nanoparticle at $r < a$ is given by a MD mode profile
with the $y$-aligned magnetic moment:
\begin{equation} \label{eq1}
\vec E_{M}^{\om}(\vec r) = E_0 \frac{3 i c_1 }{2} \vec{M}^{(1)}_{o11}(k_2 (\omega),\vec r) \:.
\end{equation}
This geometry corresponds to $\ve m || \ve e_y$ or $\theta=\pi/2$, $\varphi=\pi/2$ in Fig.~\ref{fig:MDpump}.
 Integration of the  trigonometric functions in Eq.~(\ref{eq:Dcoeff}) over the angles shows that within the framework of single-mode MD approximation~\eqref{eq1} and crystalline axes of material oriented according to Fig.~\ref{figBTgeome}, the multipolar composition features electric octupole and magnetic quadrupole for the AlGaAs nanoparticle,  allowing us to write down the expression for the field: 
 \begin{multline}
    {\bf{E}}^{2\omega}(r>a)= E_0 \Big[ D_{N_{o23}} \vec{N}^{(3)}_{o23} (k_1(2\omega), \vec{r}) \\ + D_{M_{e02}}\vec{M}^{(3)}_{e02} (k_1(2\omega), \vec{r}) + D_{M_{e22}} \vec{M}^{(3)}_{e22} (k_1(2\omega), \vec{r}) \Big] \:.
\end{multline}
 The multipolar amplitudes $ D_{No23}$, $D_{Me02}$ and $D_{Me22} $ can be also conveniently found using the {\em Lorentz lemma} following the procedure described in Ref.~\cite{Smirnova2018}, being alternative to the Green's function integration in Section~\ref{sec:II-plane}. This allows us to write  the expression for $D$-amplitudes through transmission coefficient $t^{E,M}_n(a)$ (see Ref.~\onlinecite{Smirnova2018}) of the incident spherical wave irradiating the particle.   After some algebra, we obtain compact analytical expressions for the multipolar coefficients: 
\begin{equation} \label{eq:MD_Dcoeff}
\begin{split}
    &D_{Me02} = \frac{-36i\sqrt{30\pi}\chi^{(2)} t^M_2 (k_2a)}{5{\varepsilon_2(\omega)}^{3/2}}I_2E_0{c_1(k_2a)}^2O_{02}, \\
    & D_{Me22}=\frac{18i\sqrt{20\pi}\chi^{(2)} t^M_2 (k_2a)}{5{\varepsilon_2(\omega)}^{3/2}}I_2 E_0{c_1(k_2a)}^2O_{22} ,\\
          &  D_{No23} =\frac{72\sqrt{70\pi} \chi^{(2)}t^E_3 (k_2a)}{  35\varepsilon_2(\omega)\varepsilon_2(2\omega)}\sqrt{\frac{\varepsilon_2(2\omega)}{\varepsilon_2(\omega)}}I_2E_0{c_1(k_2a)}^2O_{23}, 
     \end{split}
\end{equation}
where 
\[ I_2(k_2a)=\int_0^{k_2 a}{j_1^2(x)j_2\left(2\sqrt{\frac{\varepsilon_2(2\omega)}{\varepsilon_2(\omega)}} x\right)x^2dx },\]
\[O_{mn} = \dfrac{1}{\sqrt{n(n+1)}} \sqrt{\dfrac{(2n+1)}{4\pi} \dfrac{(n-m)!}{(n+m)!}}\:.\]

\begin{figure} [t!]
\includegraphics[width=1\columnwidth]{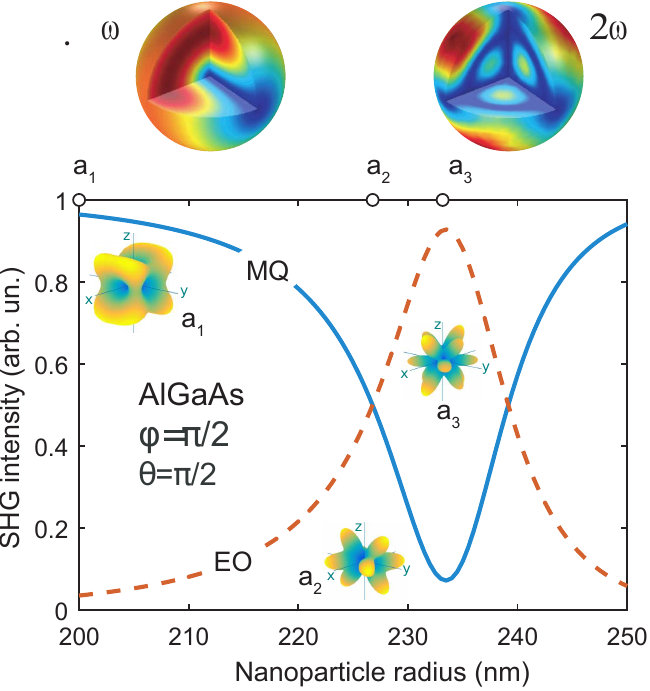} 
\caption{(Color online) SHG conversion efficiency decomposed into relative magnetic quadrupolar (blue solid line) and electric octupolar (red dashed line) contributions 
in the AlGaAs nanoparticles at the pump wavelength of 1550 nm. Insets show  \bluee{ the far-field} SH radiation patterns at $a_1=200\ \mbox{nm},\ a_2=227\ \mbox{nm}, \ a_3=234$~nm. The top figure shows the field distribution inside the nanoparticle of radius $a_2=227$ nm at the fundamental and SH frequencies. }
 \label{mlesR}
\end{figure}

\begin{figure}[t!]           

{\includegraphics[width=0.9\linewidth]{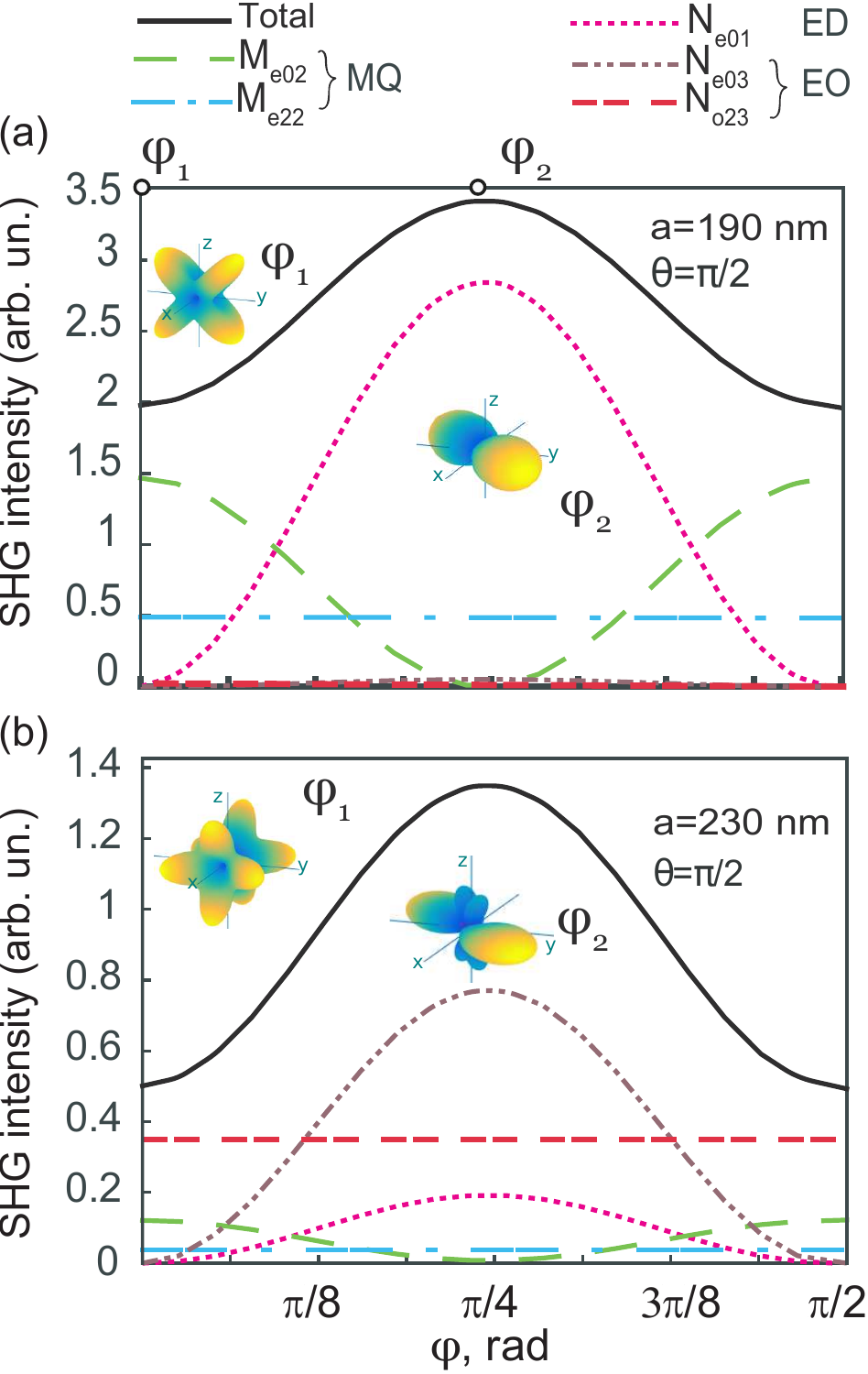}} 
\caption{Dependence of the SH intensity and generated multipoles on the pump magnetic dipole rotation by angle $\varphi$ in $xy$ plane  ($\theta=\pi/ 2 $) in AlGaAs particles of two radii $a=190$~nm (a), 230~nm (b) at the pump wavelength 1550 nm. Insets show the SH radiation patterns at $\varphi_1=0,\ \varphi_2=\pi/4$.  }\label{angle_dep_AlGaAs}   
\end{figure}

\begin{figure} [t!]
\includegraphics[width=0.8\columnwidth]{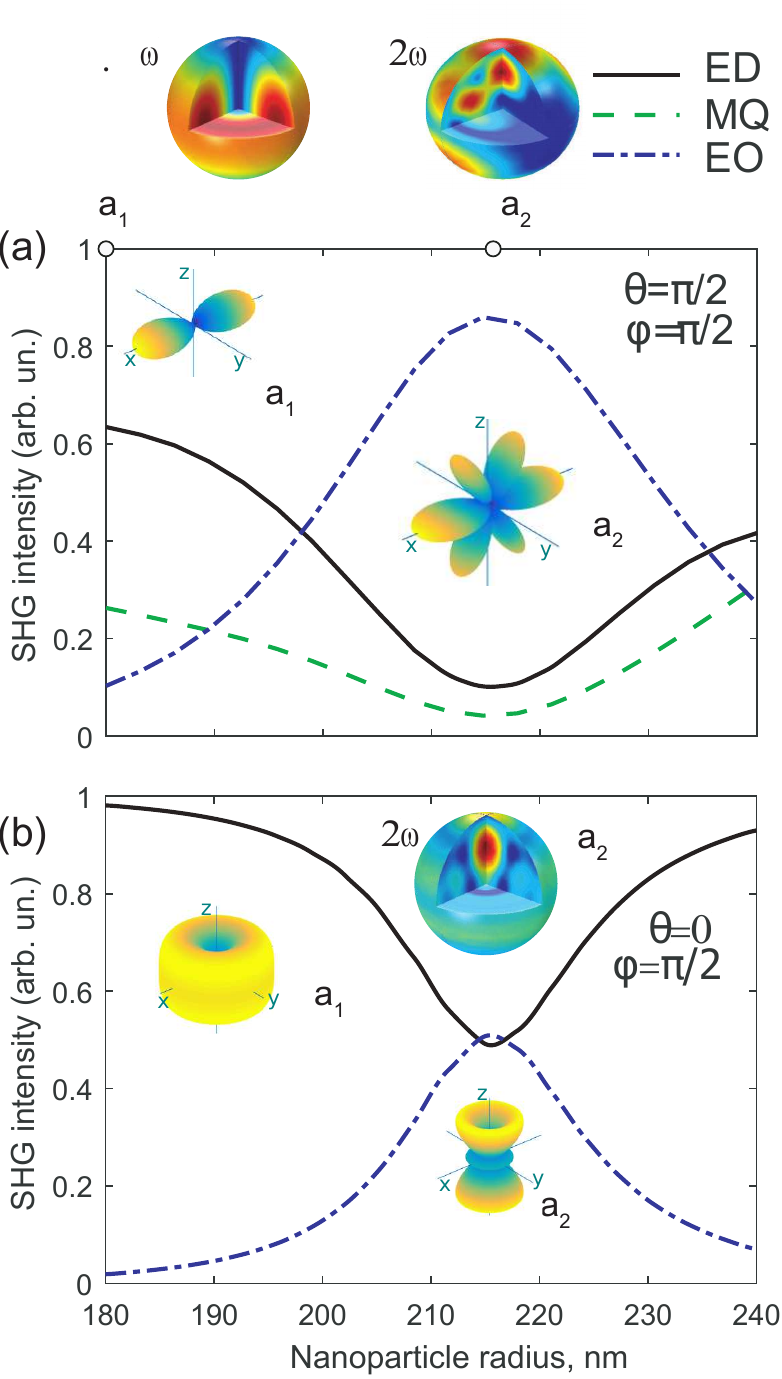} 
\caption{ Dependence of the SH intensity in BaTiO$_3$ particles  on the nanoparticle radius decomposed into relative electric dipolar (ED), magnetic quadrupolar (MQ), electric octupolar (EO)  contributions 
 at the pump magnetic dipole wavelength of 1050 nm for  $\varphi=\pi/2$,  $\theta=\pi/2$ (a) and $\theta=0$ (b). The insets show far field pattern of  SH radiation for nanoparticle sizes $a_1=180$ nm and $a_2=214$ nm. The field distribution inside nanoparticle is shown for fundamental and SH wavelength for nanoparticle radius~$a_2$.
}  \label{mlesRBT}
\end{figure}

\begin{figure}[t!]                          
\center{\includegraphics[width=0.8\linewidth]{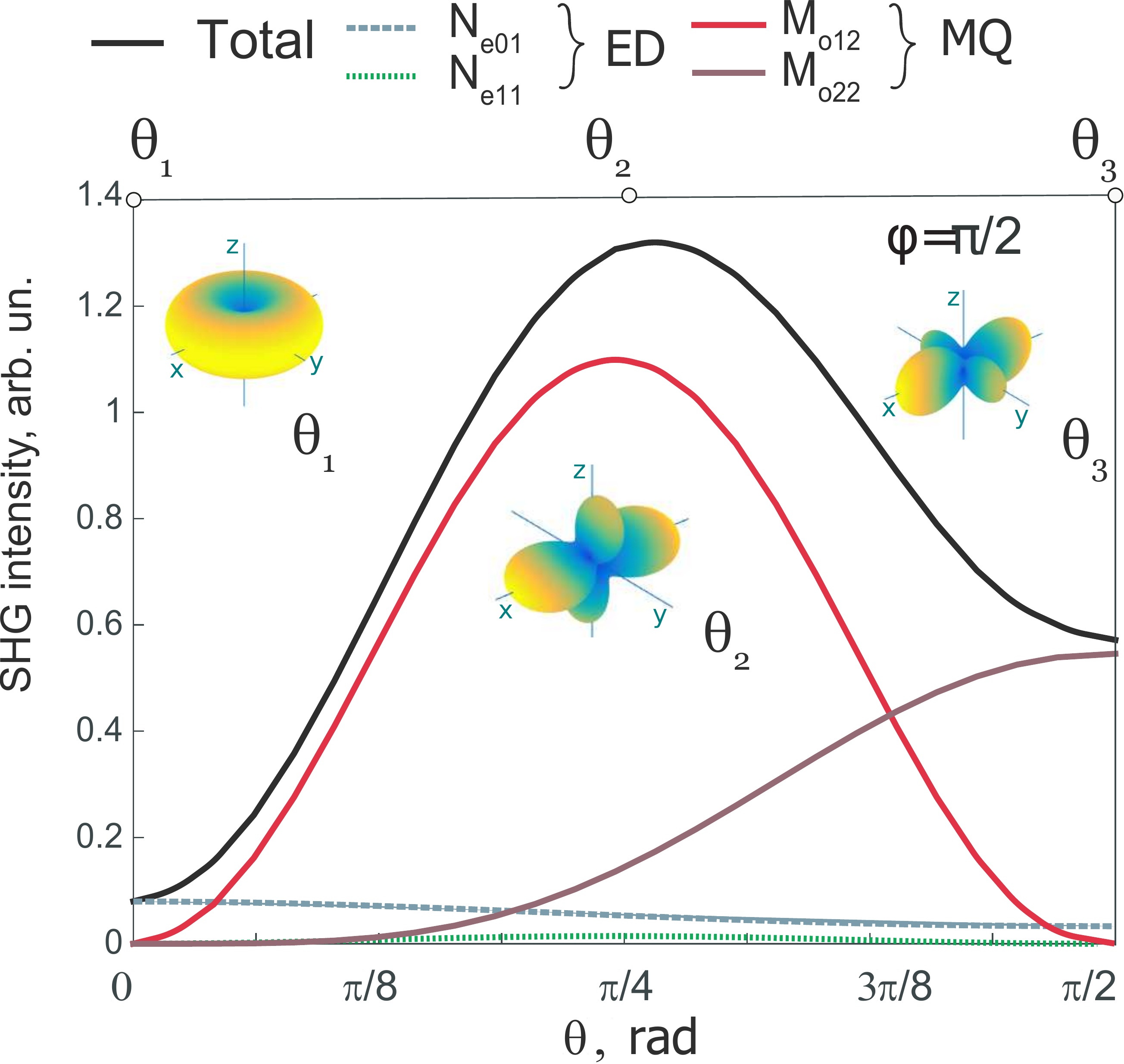}} 
\caption{Dependence of the  SH intensity and generated multipoles on the pump magnetic dipole rotation by angle $\theta$ in  the $yz$ plane ($\varphi = \pi/2$) in BaTiO$_3$ particle of radius $a$=140~nm. Insets show SH radiation patterns at $\theta_1=0,\ \theta_2=\pi/4,\theta_3=\pi/2$. } \label{angle_BT}   
\end{figure}

   The relative  contribution of different multipoles varies when the SH wavelength is tuned to corresponding  Mie resonances. This  immediately follows from the expressions for multipolar amplitudes  Eq.~(\ref{eq:MD_Dcoeff}). In Fig.~\ref{mlesR} we trace  this behaviour by plotting the dependence of the SH intensity on nanoparticle size. When  the radius is increased, the contribution of EO mode starts to dominate over the magnetic quadrupole (MQ) changing the \bluee{far-field }radiation pattern from axially-symmetric for smaller particles  to the multi-lobed pattern near the EO resonance at $a=234$~nm.  \bluee{The field distribution inside the nanoparticle (see insets in Figs.~\ref{mlesR},\ref{angle_dep_AlGaAs}, \ref{mlesRBT}) was obtained  with the help of COMSOL Multiphysics package. The radiation patterns (see insets in Figs.~\ref{mlesR},\ref{angle_dep_AlGaAs}, \ref{mlesRBT},\ref{angle_BT}), showing the distribution of the generated SH intensity in the far-field, were plotted with the use of the formula Eq.~\eqref{eq:E2omega} and were verified with COMSOL Multiphysics.}

\begin{table}[b!]
  \centering
  \caption{Single-mode excitation. Generated nonlinear multipoles in AlGaAs nanoparticle for two orientations of the pump MD moment under the rotation in the $xy$ plane ($\theta=\pi/2$). The shaded region coincides with the shaded region in Table~\ref{AlGatable}.} 
  \label{tab:table1}
\includegraphics[width=\linewidth]{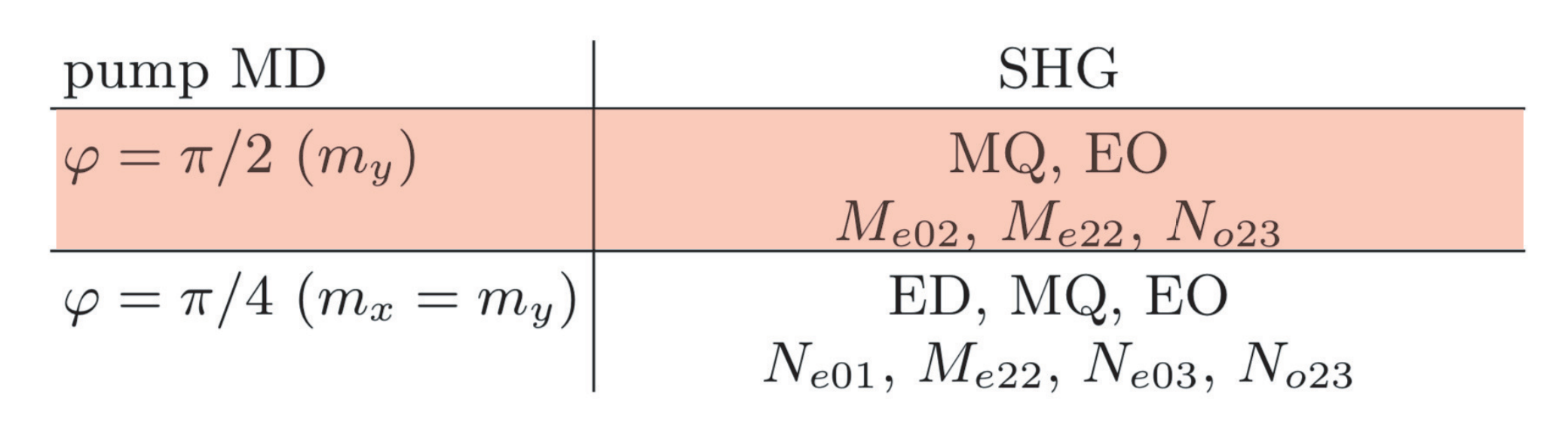}
\end{table}

\begin{table}[b!]
  \centering
  \caption{Single-mode excitation. Generated nonlinear multipoles in BaTiO$_3$ nanoparticle for two orientations of the pump MD moment under the rotation in the $yz$ plane ($\varphi = \pi/2$). The shaded region coincides with the shaded region in Table~\ref{BTztable}.}
  \label{tab:table2}
  \includegraphics[width=\linewidth]{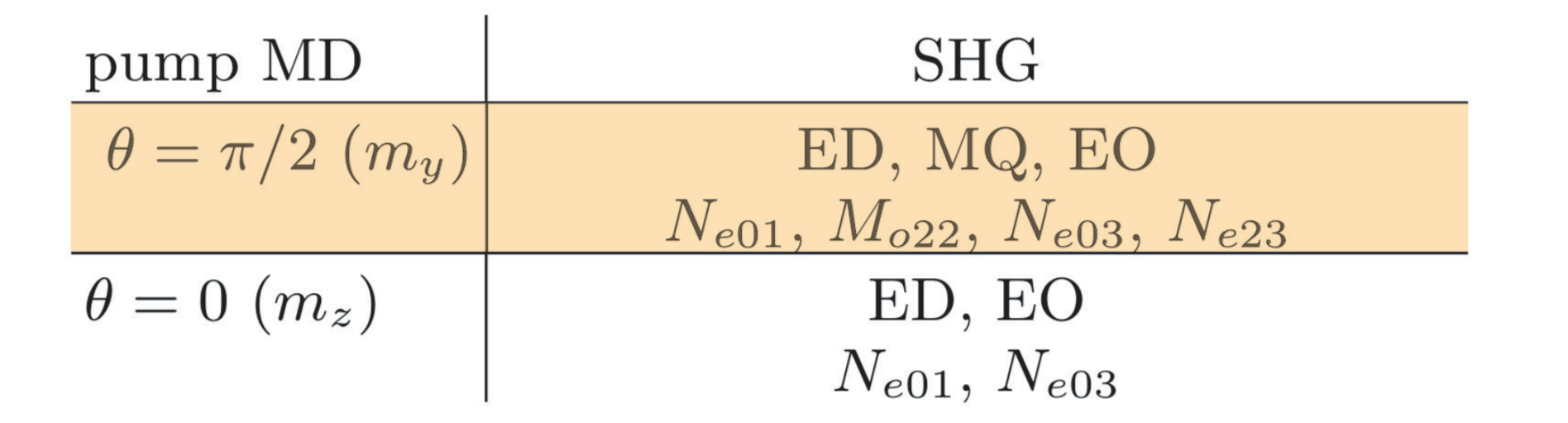}
\end{table}


 Rotation of the pump magnetic dipole in the $xy$  plane ($\theta =\pi/2$ in Fig.~\ref{fig:MDpump})  enables the generation of the ED mode in AlGaAs nanoparticle, which is also illustrated in Table~\ref{tab:table1}. The presence of the specific modes in the SH spectrum will be discussed in detail in Sec.~\ref{Sec:Discussion} basing on symmetry reasons. For the smaller nanoparticle of $a=190$ nm (Fig.~\ref{angle_dep_AlGaAs}(a)), being remote 
from the EO-resonant size, the EO contribution in the SH field is negligible. At $\varphi=0$ or $\pi/2$, 
the SH radiation is dominated by the magnetic quadrupole. If the pump magnetic dipole is rotated by $\varphi=\pi/4$, the ED relative contribution exceeds MQ and the radiation pattern significantly changes.
For the larger nanoparticle of $a=230$ nm [Fig.~\ref{angle_dep_AlGaAs}(b)], the EO term dominates in the SH emission. 

The results of analogous calculations for BaTiO$_3$ particle are shown in Fig.~\ref{mlesRBT}. Here, two different orientations of the magnetic dipole ($ \varphi = \pi/2, \  \theta=0$ and $\varphi = \pi/2, \ \theta=\pi/2$) are shown and the resonant switching between dipolar, quadrupolar, and octupolar modes is observed.

In Fig.~\ref{angle_BT} we illustrate the effect of the MD rotation in  the $yz$-plane ($\varphi=\pi/2$) for the BaTiO$_3$ particle of radius $a=140$~nm, corresponding to the MQ peak in Fig.~3(b). Rotation of MD in the $xy$-plane will not give any changes due to symmetry of \bt \ lattice with respect to this rotation. At $\theta=0$ the induced nonlinear source does not contain a MQ component, and, thus, weak SHG is determined by the non-resonant ED. When the angle $\theta$ is increased, the total SHG intensity grows and the leading contribution to the SH radiation originates  from the resonant multipole MQ. 





\section{Symmetry analysis for the second-harmonic generation}
\label{Sec:Symmetry}
In this section, we analyze the possibility of SHG through different multipole channels which is determined by the particular symmetry of modes at the fundamental and SH frequencies.
The  integrals over the nanoparticle volume
\begin{multline}
I_{\mu n,\mu' n'\to \mu'' n''}=\\
=\int\limits_{r<a} \rmd V \chi^{(2)}_{\alpha\beta\gamma}W^{(1)}_{\alpha,\mu n }(\bm r)
W^{(1)}_{\beta,\mu' n'}(\bm r)W^{(1)}_{\gamma,\mu'' n''}(\bm r)\:,\label{eq:I}
\end{multline}
determine the contributions of the multipoles $\mu' n'$,$\mu n$ to the SHG expansion coefficients 
$D_{W^e_o mn}$ in Eq.~\eqref{eq:Dcoeff}.
Here, the indices $\mu$,$\mu'$,$\mu''$ stand for the parity $e, o$ and the projection $m$  of vector spherical harmonics $\vec W^{(1)}_{^e_omn} (\vec r)$. For each particular value of indices such integrals can be readily calculated analytically, and a large number of them turn out to be zero. Our goal is to reveal the general origin of these cancellations. All our considerations are based on the following general theorem \cite{landau, Dresselhaus2008}. Let $\psi_i^{(\alpha)}$ be one of the basis functions of an irreducible (non-unit) representation $\alpha$ of a system's symmetry group. Then the integral of this function over the configuration space of the physical system vanishes identically: $\int \psi_i^{(\alpha)} dq =0 $. In order for  the integral to be non-zero, the integrand must contain a term that is invariant when any of the symmetry operations of the group are applied, otherwise the integral vanishes.

\begin{figure}[t]
\centering\includegraphics[width=0.4\textwidth]{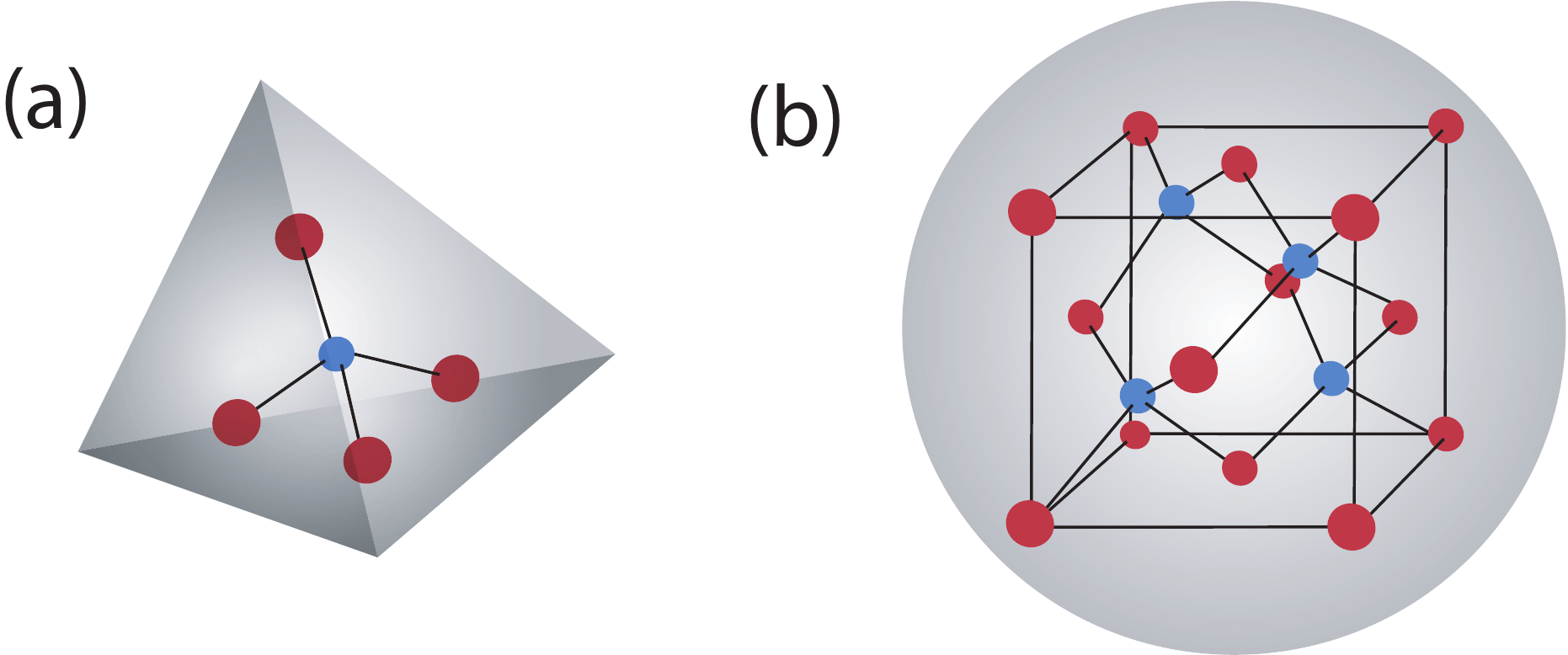}
\caption{Schematic illustration of (a) an atomic cluster 
with  $T_{d}$ symmetry
and (b) a spherical nanoparticle made of material with $T_{d}$ crystalline symmetry.
 Cation and anion atoms are shown by blue and red color, respectively.
}\label{fig:symm}
\end{figure}

 In the considered case, the cancellations stem from both (i)  the microscopic crystalline symmetry of the material and (ii)  from the macroscopic spherical symmetry of the nanoparticle as a whole. 
  In order to illustrate this distinction, we show in Fig.~\ref{fig:symm} (a) a tetrahedral nanoparticle cut from a material with $T_{d}$ point group symmetry, e.g. AlGaAs, and  in Fig.~\ref{fig:symm} (b)  the spherical nanoparticle made of the crystal with $T_{d}$ point group symmetry. Then, since tetrahedron has the same $T_d$ symmetry, all the allowed transitions will be defined by the crystalline symmetry only.   
However,  spherical symmetry of the nanoparticle in Fig.~\ref{fig:symm} (b) imposes additional restrictions on the second harmonic generation, absent in a  tetrahedron. The restrictions are provided by a certain parity of sphere eigenmodes  with respect to the symmetry transformations of a sphere.  Hence,  we can make  use of selection rules based on spatial parity of eigen modes. As such, most of the cancellations of the integrals Eq.~\eqref{eq:I} stem from simple spherical symmetry considerations, discussed in the following  Sec.~\ref{sec:spherical}. The rest of the relevant cancellations can be explained as inherited from the  crystal point group symmetry and will be considered  in Sec.~\ref{sec:crystal}.
  
We  note, that in our consideration we neglect the roughness of the spherical particle surface imposed by the crystalline structure as we assume nanoparticle containing large enough number of atoms. {  The possible nonlocal corrections to the linear dielectric response of the nanoparticle, arising from the spatial dispersion of the permittivity, and sensitive to the difference between tetradehral and spherical symmetry~\cite{agranovich2013crystal}, are also  neglected here, and the particle is described by the local isotropic permittivity tensor}.

\subsection{Restrictions imposed by the spherical symmetry}\label{sec:spherical}
\blue{The symmetry analysis of the integrals Eq.~\eqref{eq:I} becomes more straightforward when 
 the spherical harmonics $\bm W$ are presented in a vector form.  To this end we use the following relationship between the  Cartesian basis vectors and the  electric dipole harmonics in the limit $k\to 0$:}
$\bm N^{(1)}_{e11}(0, \vec r)\equiv \bm N_{x}\propto \bm e_{x}$,
$\bm N^{(1)}_{o11}(0, \vec r)\equiv \bm N_{y}\propto \bm e_{y},$
 $\bm N^{(1)}_{e01}(0, \vec r)\equiv \bm N_{z}\propto \bm e_{z}$. 
 This allows us to rewrite  the integral Eq.~\eqref{eq:I} as
\begin{multline}
I_{\mu n,\mu' n'\to \mu'' n''}\propto
  \chi^{(2)}_{\alpha\beta\gamma} \int\limits_{r<a} \rmd V
[\bm N_{\alpha}\cdot \bm W_{\mu n }(\bm r)]\\\times
[\bm N_{\beta}\cdot \bm W_{\mu' n'}(\bm r)]
[\bm N_{\gamma}\cdot \bm W_{\mu'' n''}(\bm r)]
\:.\label{eq:I2}
\end{multline}
 
Let us consider the integral \eqref{eq:I2} in more detail. First of all, it 
contains  a sum of several integrals of three scalar products of vector spherical harmonics, where each term corresponds to one of $\chi^{(2)}$-tensor components. We are seeking for the cases when the  integrand is invariant under the transformations of the $O(3)$ symmetry group. We expect the integral to be  non-zero, if at least one of the integrand terms  contains a function invariant under all rotations and inversion transformation.  
 The scalar products  entering Eq.~\eqref{eq:I2} can be readily expanded over the scalar spherical harmonics, see Refs.~\cite{DeBeer2009, VMK} and Appendix \red{\ref{scalar_prod_app}}.
Afterwards, the resulting integrals can be  analyzed for different $\chi^{(2)}$-tensor components and three general {\it Rules} $A, B,\ \text{and}\ C$, governing whether the integrals are zero or not, can be formulated for each component separately. 

\begin{figure}[!h]	\includegraphics[width=0.9\linewidth]{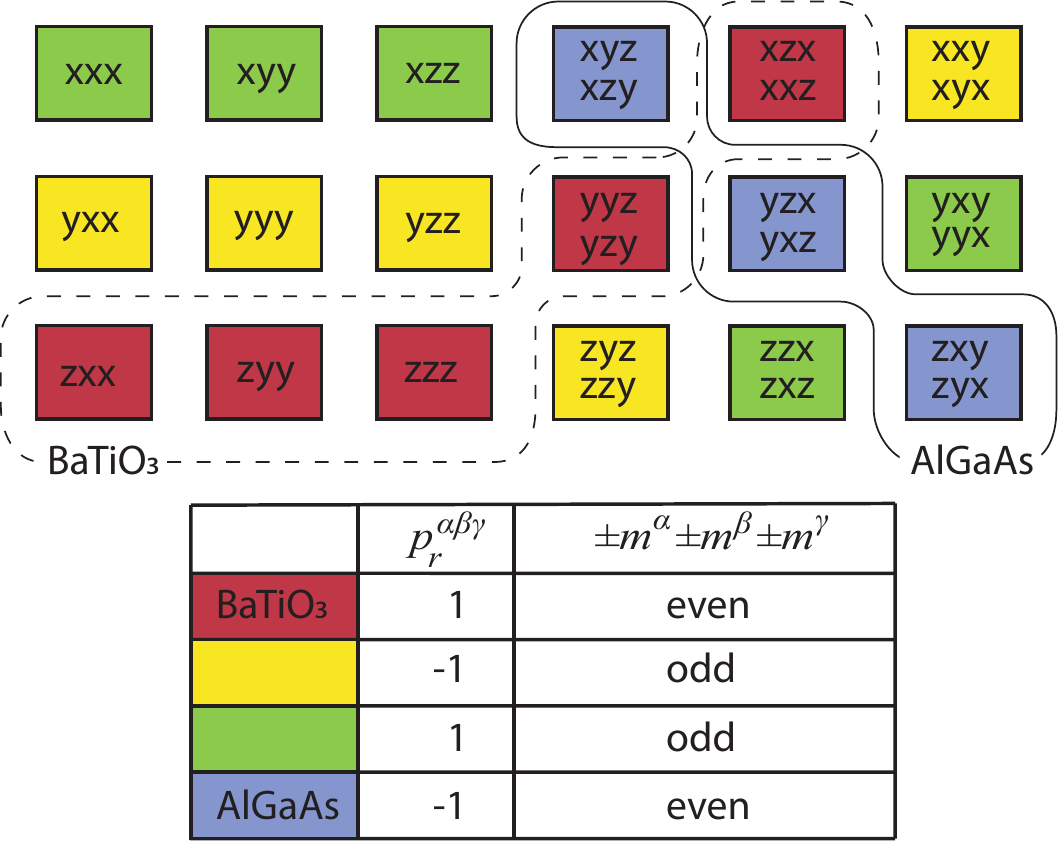}
	\caption{Parity indices of the $\hat \chi^{(2)}$-tensor components. The solid and dashed  lines show the non-zero tensor components for \ga  \ and \bt\ correspondingly for the crystal lattice oriented according to Fig.~\ref{figBTgeome}. }
	\label{tenscolor}
\end{figure}

\paragraph*{Rule A: Parity under inversion and reflection in the $y=0$ plane.} \label{parity_paragraph}
The vector spherical harmonics (Appendix~\ref{app:definitions})  are transformed in the same way as the real scalar spherical harmonics $\psi_{^e_omn}$  \cite{In2007, Stein, Theorem2008, James1976}  under the coordinate $rotations$, while under $inversion $ $\psi_{^e_omn}$ and $\bm{N}_{^e_omn}$ acquire a sign $(-1)^{n}$ and $\bm{M}_{^e_omn}$ acquires a sign $(-1)^{n+1}$, so the parity of magnetic vector harmonics is inverse to the parity of electric and scalar harmonics.
 We  introduce the {\it parity indices} $p_i=(-1)^{n}$  for $\bm{N}_{^e_o mn}$, and $p_i=(-1)^{n+1}$  for $\bm{M}_{^e_o mn}$ describing the behavior of spherical harmonics under the spatial inversion. 
Another parity index $p_{r}$ describes the behavior  of the harmonics under the reflection in the 
$y=0$ plane, equivalent to the change of the azimuthal angle $\phi\to -\phi$.
The  functions $\psi_{emn}$, $\bm{N}_{emn}$ and $\bm{M}_{omn}$ 
are even with respect to such reflection, $p_{r}=1$, while   the functions $\psi_{omn},  \bm{N}_{omn}$ and $\bm{M}_{emn}$ are odd, $p_{r}=-1$. Thus, the  inversion and reflection parity selection rules   can be summarized as 
     \begin{align}
{ p_i  p_{i'} p_{i''}=-1}&& \ \text{(inversion)},\label{eq:rule-i}\\
{p_r^{\alpha\beta\gamma} p_r   p_{r'}   p_{r''} =1}&&\text{(reflection)}\:.\label{eq:rule-r}
  \end{align} 
The rule Eq.~\eqref{eq:rule-i} { is applied to the right-hand side of  Eq.~\eqref{eq:I2} as a whole. }
The rule 
Eq.~ \eqref{eq:rule-r} is applied to the individual products of different Cartesian components corresponding to each nonzero element $\chi_{\alpha\beta\gamma}$  of the nonlinear susceptibility tensor in Eq.~\eqref{eq:I2}. The factor 
  $p_r^{\alpha\beta\gamma}$ in Eq.~\eqref{eq:rule-r}  is the parity of the product $x_{\alpha}x_{\beta}x_{\gamma}$ under the reflection, which is illustrated in Fig.~\ref{tenscolor}. {In the following, we will also use notation of $p_{i(r)}^{\om}$ or $p_{i(r)}^{2\om}$ for the parity indices corresponding to the  fundamental or the SH modes. } 

\paragraph*{Rule B: Conservation of the angular momentum projection.} 
Once the scalar products in Eq.~\eqref{eq:I2} are calculated, the  matrix element is reduced to the overlap integral of scalar spherical harmonics.  The angular momentum projection rule for the tesseral harmonics can be written as
       \begin{gather}
{\pm m^{\alpha}\pm m^{2\omega} \pm m^{\beta} \pm m'^{\omega} \pm m^{\gamma} \pm m''^{\omega}=0\:.}\label{eq:projections_sum}
  \end{gather} 
    The matrix element \eqref{eq:I2} can be non-zero only if there exists a combination of signs when  
Eq.~\eqref{eq:projections_sum}  is satisfied. 

\paragraph*{Rule C: Triangle inequality for the total angular momentum.} 
This rule can be formulated as
         \begin{gather}\label{eq:triangle}
{-h+|n'^{\omega}-n''^{\omega}|\leq n^{2\omega}\leq |n'^{\omega}+n''^{\omega}|+h}\:,
  \end{gather}
where the index $h \in [0..3] $ denotes the number of electric harmonics under the integral \eqref{eq:I}. 
The presence of the index $h$  stems from the fact that the Cartesian components of vector electric harmonics with the total angular  momentum $n$ include only the scalar harmonics with the momentum $n\pm 1$, while the projections of  magnetic harmonics include only the states with the same angular momentum $n$.

Now, let us illustrate the rules obtained above with some specific examples. We start with examining the  possibility of the generation of $z$-polarized electric dipolar mode $(\vec N_z= \vec N_{e01})$ in a AlGaAs nanoparticle by combining the $y$-polarized magnetic dipolar mode $(\vec M_y= \vec M_{o11})$  and the  $x$-polarized electric dipolar mode  $(\vec N_x= \vec N_{e11})$ . Here  $p_i  p_{i'} p_{i''}=1$, so such generation is prohibited by the first selection rule. 
Next, we try to replace $\vec N_z$ by some electric quadrupolar mode, for example, $\vec N_{o12}$. For this mode we have $p_i  p_{i'} p_{i''}=-1$,      $p_r^{\alpha\beta\gamma} =-1$ according to the Fig.~\ref{tenscolor}, and $p_r^{\alpha\beta\gamma} p_r   p_{r'}   p_{r''} = (-1) \cdot (-1) \cdot 1 \cdot 1=1$. Hence, such process is not prohibited by the first two rules.  But the sum of angular momentum projections  \eqref{eq:projections_sum} for considered three multipoles is always odd, while for the tensor component it is even (Fig.~\ref{tenscolor}). This means that the  total sum is never zero, and such generation process is prohibited by the second rule.

These rules  above provide large number of possible cancellations, however, some exceptions are possible due to the  properties of scalar products. To get all cancellations for the specific tensor component, we either should use the rules of how three scalar products are coupled, following the algorithm given in Appendix \ref{scalar_prod_app}, or apply additional symmetry reasons, discussed below.

\subsection{Restrictions imposed by the crystal point group symmetry}\label{sec:crystal}
In the previous Sec.~\ref{sec:spherical}, we have separately considered the cancellations of the 
terms in Eq.~\eqref{eq:I} corresponding to the individual components of the  $\chi^{(2)}$-tensor.
However, some of the components $\chi^{(2)}_{\alpha\beta\gamma}$ are equal due to the crystal point group symmetry, that can result in additional cancellations after the summation over tensor components is performed. Such mutual cancellations are taken care of by the theory of representations of symmetry groups in a universal automatic fashion. The detailed analysis  is given below.

\paragraph{Transformation of the matrix elements under symmetry operations }
In order to determine the behavior of the matrix elements Eq.~\eqref{eq:I} under 
the application of the point symmetry group operation, 
we consider how vector spherical harmonics are transformed. Here, one has to take into account that the 
transformed harmonic is in general expressed via a sum of the harmonics with different angular momentum projections  $m$ and parity $e/o$, but the same polarization {(M or N)} and the same total angular momentum~$n$ \cite{In2007, Stein, Theorem2008, James1976}.
The scalar products, entering the integral Eq.~\eqref{eq:I2} are transformed 
under the symmetry operation $\bm r\to D^{(1)}\bm r$ as
\begin{equation}\label{eq:rule}
\vec N_{\alpha}(D^{(1)}\vec r)\cdot\vec W_{ \mu n}(D^{(1)}\bm r)=  D^{(1)}_{ \widetilde  \alpha   \alpha } D^{(n)}_{ \widetilde  \mu \mu } \vec N_{\widetilde\alpha}(\vec r) \cdot \vec W_{\widetilde  \mu n}(\vec r)
\end{equation} 
Here  $ D^{(n)}_{\widetilde  \mu \mu}(g)$ are the representation matrices of the symmetry operation $g$ with momentum $n$ for tesseral harmonics \cite{Aubert2013}, in case of rotations they reduce to the combinations of the Wigner matrices. As an example we consider the case when the $\vec W$ harmonics are  the electric dipole harmonics, $\vec W_{\mu n}\to\vec N_{\beta}$, $\beta=x,y,z$ and $D^{(1)}$ is a rotation around the $z$ axis, e.g. $D^{(1)}_{xx}=D^{(1)}_{yy}=\cos \phi$,
$D^{(1)}_{xy}=-D^{(1)}_{yx}=-\sin\phi$, $D^{(1)}_{zz}=1$. In this case Eq.~\eqref{eq:rule} simplifies to
\begin{equation}
[\vec N_{\alpha}\cdot\vec N_{\beta}](D^{(1)}\vec r)=  D^{(1)}_{\widetilde \alpha   \alpha } D^{(1)}_{\widetilde \beta   \beta } \vec N_{\widetilde\alpha}(\vec r) \cdot \vec N_{\widetilde  \beta}(\vec r)\:,
\end{equation} 
i.e. a scalar product of two electric dipole modes is transformed as a second-rank cartesian tensor.

The condition that the  integral   Eq.~\eqref{eq:I2}  remains invariant under the symmetry transformation Eq.~\eqref{eq:rule} is written as
\begin{multline}
I_{\mu n,\mu' n'\to \mu'' n''}= D^{(n)}_{\widetilde \mu \mu }D^{(n')}_{\widetilde \mu' \mu'}
D^{(n'')}_{\widetilde \mu'' \mu''}
\widetilde I_{ \widetilde \mu n,\widetilde \mu' n'\to \widetilde \mu'' n''}\:,\label{eq:gI}
\end{multline}
where the matrix elements
$\widetilde I$ are given by Eq.~\eqref{eq:I} with
$\chi_{\alpha\beta\gamma}$ being replaced by
\begin{equation}
\widetilde \chi^{(2)}_{\alpha'\beta'\gamma'}=D^{(1)}_{\alpha\alpha'}D^{(1)}_{\beta\beta'}D^{(1)}_{\gamma\gamma'}\chi^{(2)}_{\alpha\beta\gamma}   \label{eq:chi2}
\end{equation}

\paragraph {Application to harmonic generation}
The most general consideration would require an expansion of the $\chi^{(2)}$ tensors
Eq.~\eqref{eq:chiBT}, Eq.~\eqref{eq:chiAlGaAs}  of BaTiO$_{3}$ and AlGaAs, transforming under the spherical symmetry operations according to  Eq.~\eqref{eq:chi2}, over the irreducible representations of the $O(3)$ symmetry group. However, in practice it turns out that the relevant cancellations of the matrix elements Eq.~\eqref{eq:I2},
not captured in the previous Sec.~\ref{sec:spherical}, can be explained
in a simpler way. Instead of the whole $O(3)$ group it suffices to  apply a crystal point subgroup of the  $O(3)$ group, i.e. to use a smaller set of symmetry operations. When the crystal point group operation is applied, the tensor $\chi^{(2)}$  stays invariant, which means that  $\widetilde\chi=\chi$,
and $\widetilde I=I$. Hence, the matrix elements $I$ in Eq.~\eqref{eq:gI} are transformed as a direct product of the three  representations governing the transformation of the corresponding vector spherical harmonics.
In order to stay non-zero, the integrals Eq.~\eqref{eq:gI} should contain a combination, invariant to the symmetry operation of the crystal. More formally, the reducible representation governing the transformation  Eq.~\eqref{eq:I2} should contain an identity representation.

The symmetry analysis of the second harmonic generation is then reduced to 
(i) expanding the vector spherical harmonics over the basis functions of the irreducible representation of the crystal point symmetry group and (ii) using the  Clebsh-Gordan coefficients available  for all point groups~\cite{Ivchenko1995,  Dresselhaus2008}. 
The classification of the dipole, quadrupole and octupole  spherical harmonics  for $T_{d}$ and $C_{4v}$ point groups of AlGaAs and BaTiO$_{3}$, respectively, is given in Appendix~\ref{app:hrmncs_symm_table} and  Table~\ref{table:symm_irr_reps}. \bluee{The examples of basis functions of irreducible representations, which behave in the same way under symmetry transformations as the spherical harmonics, are also given}.
The expansion has been done using the transformation properties of vector spherical harmonics.
  For example, we see from Table~\ref{table:symm_irr_reps}  that the electric dipole modes are transformed in the $T_{d}$ group according to the $F_{2}$ irreducible  representation, i.e. as  components of the radius-vector $\bm r$. Conversely, the magnetic dipole modes behave as pseudo-vector components, i.e. according to the $F_{1}$ representation.
  
Such  approach allows  us to find all the selection rules of the nanoparticle with the same or higher symmetry as the   material (Fig.~\ref{fig:symm}a)). It can be a tetrahedral nanoparticle for $T_{d}$ or a quadrangular pyramid nanoparticle for $C_{4v}$. This is possible due to the fact that we consider the integrand behavior under such transformations  only. The  further cancellations can appear due to the scalar product properties, for example, $z$-component of $\vec M_{e01}$ harmonic is zero, so it can provide some extra restrictions (see Appendix \ref{scalar_prod_app} ).



\section{Discussions}
\label{Sec:Discussion}

 

Let us apply the developed selection rules to the cases studied  in Sec.~\ref{sec:II-plane+single}, where we already discussed the absence of particular harmonics in the generated field.

{\it Plane-wave excitation}. Under the excitation of \bt \ (\ga) nanoparticle with a plane wave, we observed the absence MD (ED) modes in the SH field. It is illustrative to start with the restrictions imposed by the spherical symmetry of modes that account for the most of the selection rules.

Here, we will consider only the dipole terms in the excitation, and higher modes can be treated analogously. Applying Rule A {to the \bt \ nanoparticle} in the case of  dipole modes ($n=1$) we do not obtain any restrictions due to the inversion parity, { because we  both terms $ED \otimes ED$ and $MD \otimes ED$ are contained in the  fundamental field}. One  can find  that the reflection parity of the SH mode should be $p_r^{2\om}=1$.   Indeed,   according to the Mie theory generated dipole modes at the fundamental frequency can be only  $\ve N_{e11}$ and $\ve M_{o11}$, and for both of them $p_r^{\omega}=1$, while for \bt \ $\hat \chi^{(2)}$ tensor $p_r^{\alpha\beta\gamma}=1$  (see Fig.~\ref{tenscolor}).  From the  angular momentum projection conservation Rule B, we find the  limits for angular momentum of SH modes. For the tensor components the sum ${\pm m^{\alpha} \pm m^{\beta}  \pm m^{\gamma}}$ is even according to Fig.~\ref{tenscolor}. Thus, ${\pm m^{2\omega} \pm m'^{\omega} \pm m''^{\omega}}$ should also be even, and from the Mie theory it follows that  $m'^{\omega}=m''^{\omega}=1$, which makes  $m^{2\omega}$ to be even. This immediately rules out all magnetic dipole modes as the only  dipole mode with even $m$ and $p_r^{2\omega}=1$ is $\ve M_{o01}$, which is identical to zero. Electric dipole mode $\ve N_{e01}$ has the same reflection parity and, thus, is allowed in the SHG process (see Table~\ref{BTztable}).  These reasons also show that the higher order magnetic and electric modes can also exist. According to Rule C the highest possible harmonic generated from the dipole modes is the electric mode with $n=5$ as shown in Table~\ref{BTztable}: $ED\otimes ED\rightarrow ET$.

\begin{table}

\begin{center}
\caption{Possible multipoles generated in second harmonic by the dipolar terms products in the incident field. BaTiO$_3$ lattice orientation is $[100] \|  \ve{e}_x,\ [010] \| \ve{e}_y, \ [001] \| \ve{e}_z$}
\includegraphics[width=\linewidth]{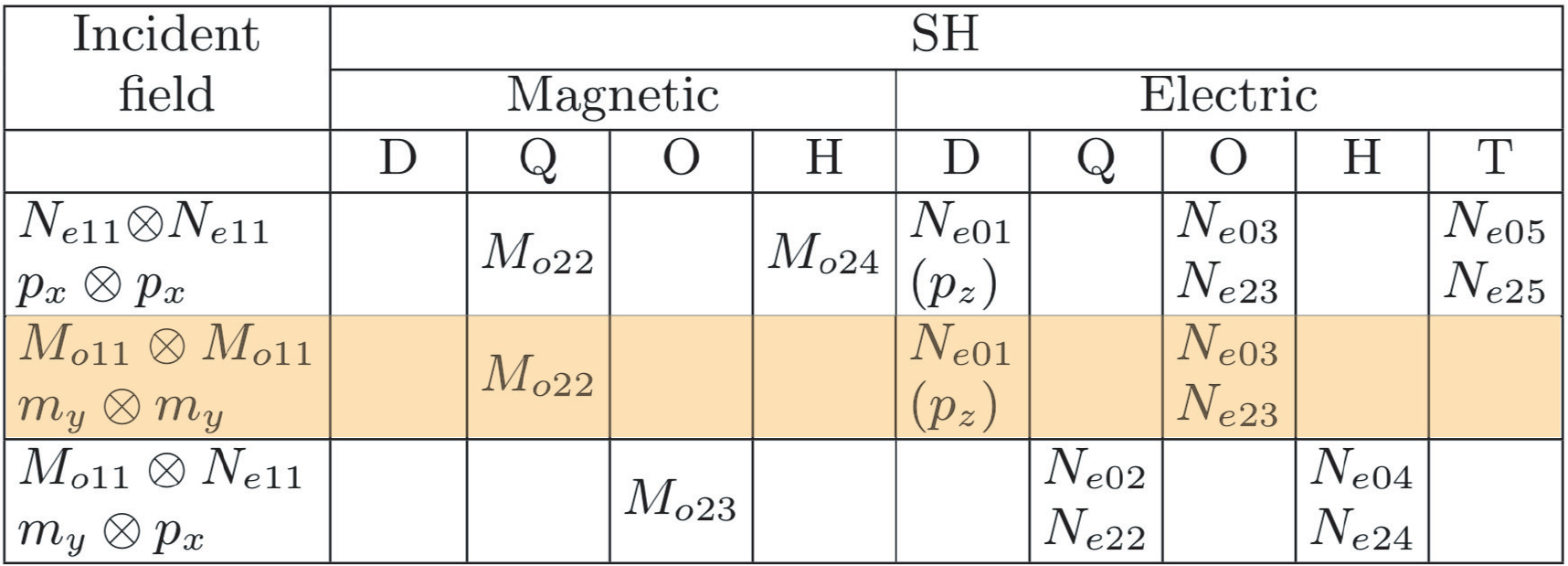}
\label{BTztable}
\end{center}
\end{table}

\begin{table}

\begin{center}
\caption{ Possible multipoles generated in second harmonic by the dipolar terms products in the incident field. BaTiO$_3$ lattice orientation is $[100] \|  \ve{e}_y,\ [010] \| \ve{e}_z, \ [001] \| \ve{e}_x$}
\includegraphics[width=\linewidth]{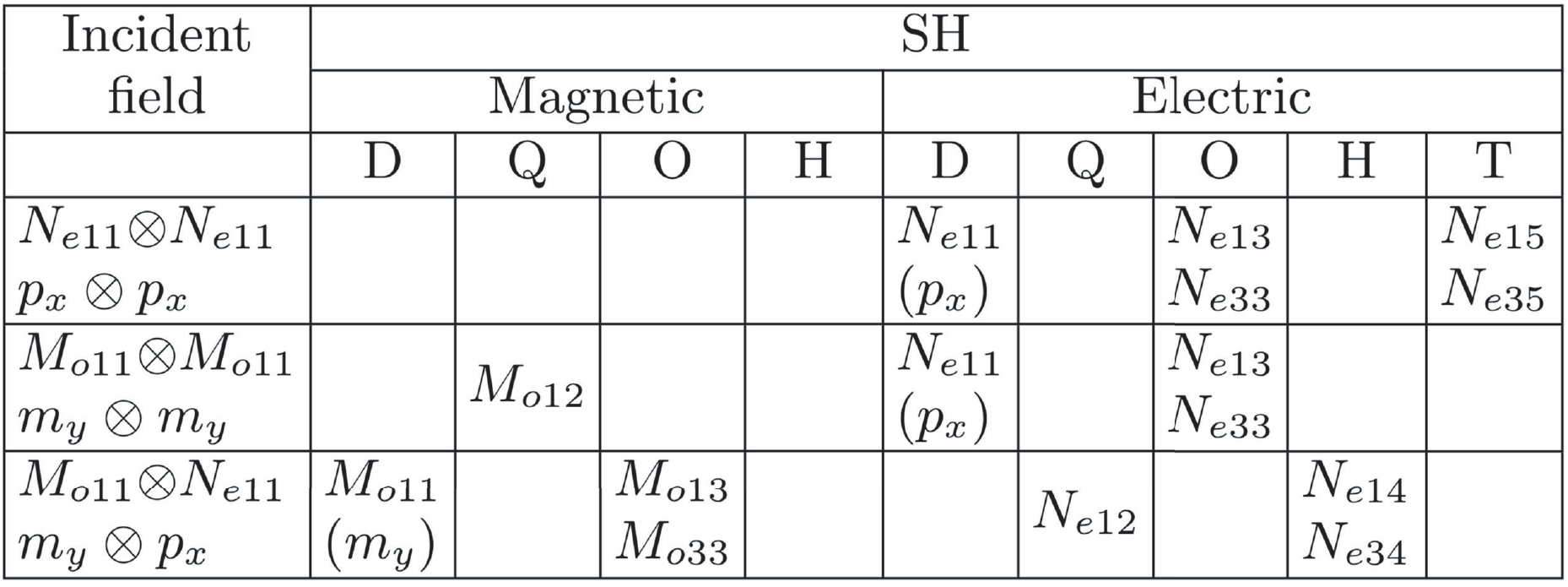}
\label{BTxtable}
\end{center}
\end{table}

The selection rules for AlGaAs are shown in the Table~\ref{AlGatable}. The same arguments as in the case of \bt \ can be applied, while considering dipole channels of SHG. The only difference is that the parity of tensor components for  AlGaAs $p_r^{\alpha\beta\gamma}=-1$ (see Fig.~\ref{tenscolor}). Rule A will be fulfilled if the parity $p_r$ of one of the modes  will be changed in sign, allowing $MD \otimes ED\rightarrow MD $ transition. Another possible channels $ ED \otimes ED \not\to ED$  or $ MD \otimes MD \not\to ED$ are forbidden, as it would require generation of $\ve N_{o01}$ mode, which is identical to zero.  Moreover, the dipole modes generation in SH field  will be still  prohibited even if the higher order modes will be excited at fundamental wavelength.




\begin{table}

\begin{center}
\caption{ Possible multipoles generated in the second harmonic by the dipolar terms products in the incident field. AlGaAs lattice rotation angle is $\beta=0^\circ$.}
\includegraphics[width=\linewidth]{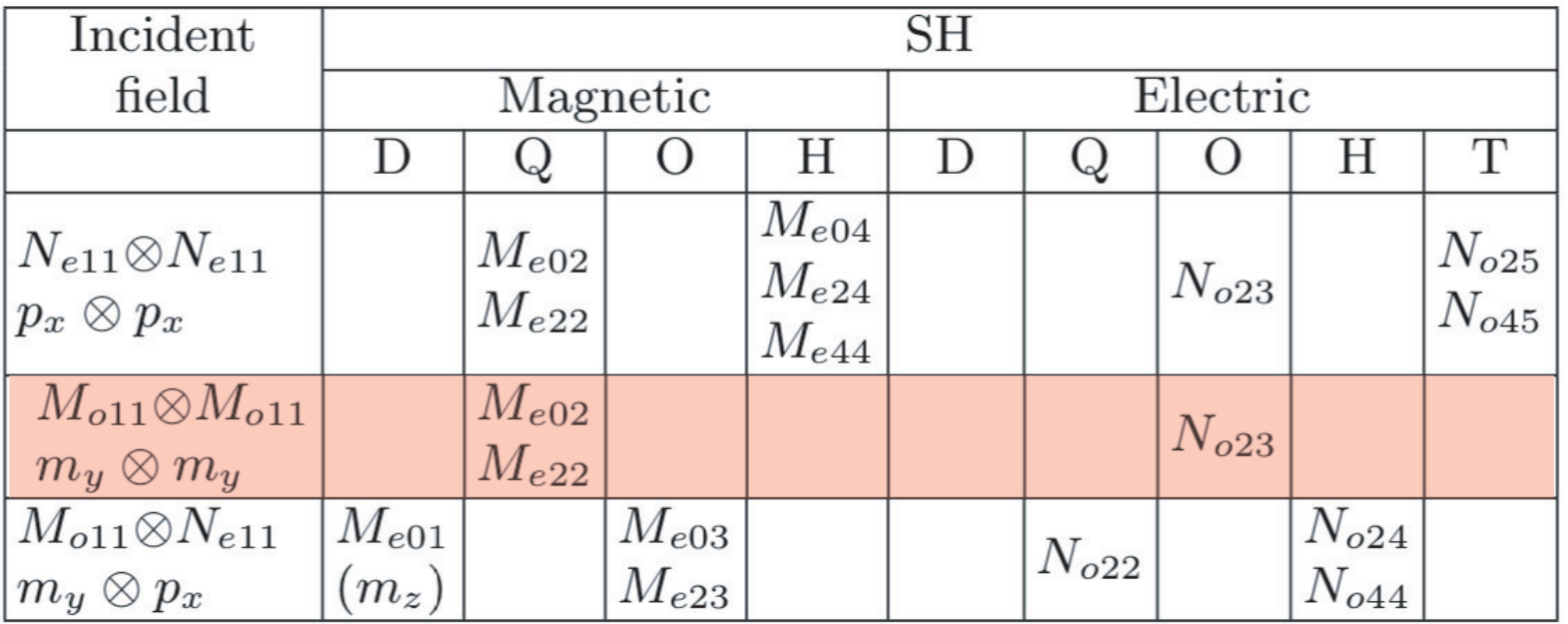}
\label{AlGatable}
\end{center}
\end{table}

{\it Single magnetic mode excitation.} In Sec.~\ref{Sec:Single_mode}, we have discussed the excitation of the SH mode with a single magnetic mode at the fundamental frequency. {Let us first study in more detail  the case of $y$-oriented dipole $\vec M_{o11}$}. Applying derived selection rules, one can get from Rule C that the highest possible generated mode is the octupole mode $n\le~3$. For \bt\ nanoparticle from Rule B, we have already established that $m^{2\om}$ should be even. The inversion and reflection parities from Rule A gives us that $p_i^{2\om}=-1$ and $p_r^{2\om}=1$. For total angular momentum value of $n=3$ this means that only electric modes should be generated (inversion rule), and they should be even (reflection rule),  which gives us for even $m$ only two possible modes: $\ve N_{e03}$ and  $\ve N_{e23}$. For $n=2$ only odd magnetic quadrupole mode possesses required reflection and inversion parity, which gives us only $\ve M_{o22}$ contribution as $\ve M_{o02}\equiv 0$. These selection rules are summarized in the highlighted row of Table~\ref{tab:table2}, which also corresponds to highlighted region of the extended Table~\ref{BTztable}. 

Until now we have considered only one orientation of the crystalline lattice, shown in Fig.~\ref{figBTgeome}. However,  the  different orientation of the \bt \ lattice provides other selection rules. For instance,  one can rotate  the BaTiO$_3$ lattice by 90$^{\circ}$ obtaining  $[100] \|  \ve{e}_y,\ [010] \| \ve{e}_z, \ [001] \| \ve{e}_x$, which changes the parity indices $p_r^{\alpha \beta \gamma}=1$, and  ${\pm m^{\alpha} \pm m^{\beta}  \pm m^{\gamma}}$ should be odd. The resulting selection rules are summarized in  Table~\ref{BTxtable}. One can see that harmonics with different projection $m$ can be generated, allowing  both MD and ED channels. This provides us an opportunity to control the SH signal by changing the relative  orientation between the field polarization and the crystalline lattice.

 The same approach allows us to analyze the modes generated at SH by pumping at single MD mode in AlGaAs nanoparticle, as shown in Table~\ref{tab:table1}. The shaded region describes excitation of the MD mode corresponding to $y$-oriented  magnetic dipole ($m_y$). The same argument as in the case of \bt\ allows up to  octupole mode generation.  We have already discussed in this section that the generation of ED mode is possible in AlGaAs due to the $p_r^{\alpha \beta \gamma}=-1$ parity. Because of that, the inversion and reflection parity values should be equal to  $p_i^{2\om}=-1$ and $p_r^{2\om}=-1$. Thus, for $n=3$  $\ve N_{o23}$ is the only nonzero mode satisfying  the parity conditions, while for $n=2$ the even magnetic modes have proper parity, thus, $\ve M_{e02}$ and $\ve M_{e22}$ are present (shaded rows in Tables~\ref{tab:table1} and \ref{AlGatable}). 
 
{\it The selection rules provided by the crystalline lattice symmetry.} So far we have discussed the selection rules which were governed by the symmetry of the vector spherical harmonics and related conditions {\it A, B,} and {\it C}. However, there are SHG channels, which are allowed by the mode symmetry, but become restricted due to crystalline symmetry only.

 For the  SHG by rotated MD in AlGaAs nanoparticle for $\phi=\pi/4$ (see  Table~\ref{tab:table1}) modes $\ve M_{o11}$ and $\ve M_{e11}$ are present in the fundamental field, while the channels of  generation of higher quadrupole modes are forbidden   $\ve M_{o11}\otimes \ve M_{e11}\not\to \ve M_{o12}, \bm M_{e12}, \bm M_{o22}$, {however}  some of them satisfy the mode symmetry rules. These processes are restricted by the crystalline symmetry rules discussed in Sec.~\ref{sec:crystal}: in the $T_{d}$ symmetry group the magnetic dipole modes are transformed as  components of the pseudo-vector $\bm L$ ($F_{1}$ representation).The modes $\bm M_{o12}$, $\bm M_{e12}$, $\bm M_{o22}$  also behave under the symmetry operations as components of a pseudo-vector $\bm L$ ($F_{1}$ representation), see the Table~\ref{table:symm_irr_reps}. Now, the physical question, of whether the SHG process is possible, is  reduced to the mathematical question of whether the direct product $F_{1}\otimes F_{1}$ contains $F_{1}$.  \bluee{The 9-dimensional reducible representation   $F_{1}\otimes F_{1}$ is  equal to a direct sum  of irreducible representations $ A_1\oplus E \oplus F_2\oplus F_1$  \cite{Ivchenko1995,Dresselhaus2008}.  If we label the 3-pseudovectors, forming the basis of  the representation $\bm F_{1}$ as $\bm M$ and $\bm M'$, the nine linear combinations transforming to 
 $ A_1$, $E$, $F_1$, $F_2$   are 
 $\bm M\cdot\bm M'$ ($A_{1}$), $\sqrt{3}(\vec M_{x} \vec M'_{x}- \vec M_{y} \vec M'_{y})$ and $ 2\vec  M_{z}\vec M'_{z}-\vec  M_{x}\vec  M'_{x}-\vec  M_{y}\vec  M'_{y}$ ($E$),
 $\vec M_{x}\vec  M'_{y}+\vec  M_{y}\vec  M'_{x}$ and two cyclic permutations ($F_{2}$) and $\bm M\times\bm M'$ ($F_{1}$).  \red{The latter must be understood componentwise.}
 We are interested only in the magnetic dipole contribution, i.e. in the pseudovector  $\bm M\times\bm M'$ transforming according to $F_{1}$. } However,  in our case $\bm M$ and $\bm M'$ are  equal, since  both modes at the first harmonic frequency belong to the same electric field. As such, the vector product $\bm M \times \bm M'$ is zero and the conversion from the magnetic dipole modes to the  $\bm M_{o12}$,$\bm M_{e12}$,$\bm M_{o22}$  quadrupolar modes is forbidden by the $T_{d}$ crystal symmetry. For the similar reasons, such process is prohibited for the conversion from the electric  dipole modes  $\ve N_{o11}$ and $\ve N_{e11}$. On the other hand, a sum frequency generation process, when the incident modes have different frequency, can be possible, since $\bm M \times \bm M'$ in general is not zero.

\bluee{
{\it Anisotropy of the linear response of the nanoparticle}. 
Finally, let us briefly discuss the effect of the uniaxial symmetry of the linear response of the dielectric tensor, present for BaTiO$_{3}$  on the obtained selection rules for the second harmonic generation.  The impact  of anistotropy on linear scattering  has been studied in details in Ref.~\onlinecite{Geng2004, Stout:06}. However, for the considered range of parameters the anisotropy is not very strong, namely $\sqrt{\eps_{xx}}=\sqrt {\eps_{yy}}=2.33$ and $\sqrt {\eps_{zz}}=2.29$ at the fundamental  wavelength of  1050~nm~\cite{Palik2012} and $\sqrt{\eps_{xx}}=\sqrt{\eps_{yy}}=2.48$ and $\sqrt{\eps_{zz}}=2.42$ at the SH  wavelength of  525~nm~\cite{wemple1}, a rigorous extension of the approach presented in this paper to the anisotropic case manifests itself a complicated problem. Thus, we have appleid numerical simulation method in order to check the  effects of the present material anisotropy on the SH field multipole content. The spectral dependence of the SHG cross section is shown in Fig.~\ref{fig:aniso} similarly to Fig.~\ref{btmult} but with account for anisotropic permittivity tensor. Simulations results have shown that the difference in the nonlinear response is rather weak, and the multipolar contents is generally preserved for the given set of the parameters.     

The further speculations on the influence of the anisotropy will bring us to the conclusion on the selection rules modification. In general,  the linear scattering of a plane wave on an isotropic particle preserves  both the multipole order $(n, m)$, and the electric or magnetic parity, namely each spherical harmonic contained in a plane wave generates a scattered  harmonic with the same  numbers $(n, m)$,  and the same electric/magnetic parity.  In the process of anisotropic particle scattering the mode numbers are not conserved \cite{Stout:06, Hui_Chen} and multipole orders $n$, their projections $m$ \red{(in case of arbitrary orientation of the optic axis of the crystal)}, and electric and magnetic degrees of freedom are getting mixed. The SHG process will \red{also} entangle the multipole orders due to the structure of the anisotropic Green's function. These two processes change the selection rules A-C. However, the formulation of the exact  selection rules in this case is a matter of the future work, there will be a particular  resemblance to the case of the SHG from a cylindrical nanoparticle with isotropic linear permittivity tensor \cite{Frizyuk}. }


\section{Conclusion}

In conclusion, we have theoretically analyzed the second harmonic generation by spherical dielectric nanoparticles made of materials with non-zero bulk second order nonlinear susceptibility tensor $\hat \chi^{(2)}$. Considering two typical crystalline solids, \bt \ and \ga, we have studied the intensity of SHG under a plane wave illumination and analyzed the contribution of different multipole components into the total SH intensity. We have shown that under the resonant excitation of a single magnetic dipole mode one can achieve control of the directionality of SH emission by rotating the  dipole moment with respect to the material's crystalline lattice. Finally, we have developed a symmetry approach which provides an  explanation why only particular modes can be observed in the SH field and defined general  selection rules for SHG. Our symmetry analysis fully agrees with numerical and analytical results, and also demonstrates promising predictive power, which can be used for design of efficient nonlinear light sources based on nanoparticle ensembles.

\acknowledgements
The authors acknowledge useful discussions with  S.E. Derkachev,  A. A. Nikolaeva, M.O. Nestoklon, E.L. Ivchenko, A.I. Smirnov. The work was supported by the Russian Foundation for Basic Research (Grant No. 18-02-00381, 18-02-01206). Numerical modeling, performed by D.S., was supported by the Russian Science Foundation (Grant No. 17-12-01574).
A.P. and M.P. have been supported by the Foundation for the Advancement
of Theoretical Physics and Mathematics ``Basis''.

\bibliography{liter_2}

\newpage
\appendix 

 
\begin{widetext}

\section{Vector spherical harmonics} \label{app:definitions}
Vector spherical harmonics used above  are defined as  
  \begin{gather}
 \vec{M}_{^e_omn}=\nabla \times (\vec r \psi_{^e_omn})\:, \\
  \vec{N}_{^e_omn}=\frac{\nabla \times \vec{M}_{^e_omn}}{k}\:,
\end{gather}
where
 \begin{gather}
\psi_{emn}=\cos m\phi P_n^m(\cos\theta)z_n(\rho)\:, \\
\psi_{omn}=\sin m\phi P_n^m(\cos\theta)z_n(\rho)
 \end{gather}
 are the scalar spherical functions, proportional to the tesseral spherical functions. Functions $z_n(\rho)$ can be replaced by  spherical Bessel functions of any type, \blue{$\rho$ is $k(\omega)r={\omega}\sqrt{\varepsilon}r/c$}.
 
\begin{gather}
{\bf \vec{M}}_{emn}(k, \vec r) = \frac {-m}{\sin(\theta)}\sin(m\phi)P_n^m(\cos(\theta))z_n(\rho){\bf \vec e_{\theta}}- \nonumber \\
-\cos(m\phi) \frac{dP_n^m(\cos(\theta))}{d\theta}z_n(\rho){\bf \vec e_{\phi}}\:,
\end{gather}
\begin{gather}
{\bf \vec{M}}_{omn}(k, \vec r) = \frac {m}{\sin(\theta)}\cos(m\phi)P_n^m(\cos(\theta))z_n(\rho){\bf\vec e_{\theta}}- \nonumber\\
-\sin(m\phi) \frac{dP_n^m(\cos(\theta))}{d\theta}z_n(\rho){\bf\vec e_{\phi}}\:,
\end{gather}
\begin{gather}
{\bf \vec{N}}_{emn}(k, \vec r) = \frac {z_n(\rho)}{\rho}\cos(m\phi)n(n+1)P_n^m(\cos(\theta)){\bf\vec e_{r}}+ \nonumber\\
+\cos(m\phi) \frac{dP_n^m(\cos(\theta))}{d\theta}\frac1{\rho}\frac{d}{d\rho}[\rho z_n(\rho)]{\bf\vec e_{\theta}}-\\-
m\sin(m\phi) \frac{P_n^m(\cos(\theta))}{\sin(\theta)}\frac1{\rho}\frac{d}{d\rho}[\rho z_n(\rho)]{\bf\vec e_{\phi}} \:,\nonumber
\end{gather}
\begin{gather}
{\bf \vec{N}}_{omn}(k, \vec r) = \frac {z_n(\rho)}{\rho}\sin(m\phi)n(n+1)P_n^m(\cos(\theta)){\bf\vec e_{r}}+ \nonumber \\+
\sin(m\phi) \frac{dP_n^m(\cos(\theta))}{d\theta}\frac1{\rho}\frac{d}{d\rho}[\rho z_n(\rho)]{\bf\vec e_{\theta}}+\\+
m\cos(m\phi) \frac{P_n^m(\cos(\theta))}{\sin(\theta)}\frac1{\rho}\frac{d}{d\rho}[\rho z_n(\rho)]{\bf\vec e_{\phi}}\:.\nonumber
\end{gather}
The Mie coefficients for the field inside the nanoparticle can be expressed as follows: 
  \begin{gather} \label{micoef}
 c_n(\om) = \frac {\left[ \rho_1 h_n(\rho_1)\right]'j_n(\rho_1) - \left[ \rho_1 j_n(\rho_1)\right]'h_n(\rho_1)}{\left[ \rho_1 h_n(\rho_1)\right]'j_n(\rho_2) - \left[ \rho_2 j_n(\rho_2)\right]'h_n(\rho_1)} \\
 d_n(\om) = \frac {\sqrt{\varepsilon_2}\left[ \rho_1 h_n(\rho_1)\right]'j_n(\rho_1) - \sqrt{\varepsilon_2} \left[ \rho_1 j_n(\rho_1)\right]'h_n(\rho_1)}{\varepsilon_2\left[ \rho_1 h_n(\rho_1)\right]'j_n(\rho_2) - \left[ \rho_2 j_n(\rho_2)\right]'h_n(\rho_1)},
\end{gather} 
\blue{here $\rho_1$ is $k_1(\omega) a={\omega}\sqrt{\varepsilon_1}a/c$, $\rho_2$ is $k_2(\omega) a={\omega}\sqrt{\varepsilon_2}a/c$}.

The Green function for a dielectric sphere of radius $a$ for $r > a > r'$ at frequency $2\omega$ is given by 

\blue{\begin{multline}
\te {{\bf G}}({\ve r, \ve r', 2\omega})= \frac {i k_2 (2\omega)}{4 \pi } \sum_{n=1}^{\infty} \sum_{m=0}^n (2-\delta_0) \frac {2n+1}{n(n+1)} \frac {(n-m)!}{(n+m)!} 
\cdot \\
\cdot  \Bigl( a_n^{(2)}(2\omega) (\vec{M}_{e mn}^{(3)}[k_1(2\omega), \ve r] \otimes{\vec{M}}^{(1)}_{e mn}[k_2(2\omega), \ve r'] +\vec{M}_{omn}^{(3)}[k_1(2\omega), \ve r] \otimes{\vec{M}}^{(1)}_{o mn}[k_2(2\omega), \ve r'])+\\
+b_n^{(2)}(2\omega)({\vec{N}}_{e mn}^{(3)}[k_1(2\omega), \ve r] \otimes{\vec{N}}^{(1)}_{emn}[k_2(2\omega), \ve r']+\vec{N}_{omn}^{(3)}[k_1(2\omega), \ve r] \otimes{\vec{N}}^{(1)}_{o mn}[k_2(2\omega), \ve r'])\Bigr) \label{gfg} \:, 
\end{multline}
where superscripts $(1)$ and $(3)$ appear, when we replace $z_n(\rho)$ by spherical Bessel functions, and  the spherical Hankel functions of the first kind, respectively, $\delta_0=1$ when $m=0$, and $\delta_0=0$ when $ m \neq 0$.}  Coefficients in the Green function have a similar denominator as the Mie coefficients:
\begin{gather}
 a_n^{(2)}(\om) = \frac {\left[ \rho_2 j_n(\rho_2)\right]'h_n(\rho_2) - \left[ \rho_2 h_n(\rho_2)\right]'j_n(\rho_2)}{\left[ \rho_2 j_n(\rho_2)\right]'h_n(\rho_1) -\mu_2/\mu_1 \left[ \rho_1 h_n(\rho_1)\right]'j_n(\rho_2)}, \nonumber \\
 b_n^{(2)}(\om) = \frac {\left[ \rho_2 j_n(\rho_2)\right]'h_n(\rho_2) - \left[ \rho_2 h_n(\rho_2)\right]'j_n(\rho_2)}{\mu_2/(\sqrt\varepsilon_2\mu_1)\left[ \rho_2 j_n(\rho_2)\right]'h_n(\rho_1) - \sqrt\varepsilon_2\left[ \rho_1 h_n(\rho_1)\right]'j_n(\rho_2)}. \nonumber
\end{gather}

The rigorous expression for the $D$-coefficients  is as follows:  
$$
D_{W_{^e_o mn}}=\left({\dfrac{2\omega}{c}}\right)^2 \frac {i k_{2}(2\omega)} {4 \pi }  (2-\delta_0) \frac {2n+1}{n(n+1)} \frac {(n-m)!}{(n+m)!}
\left(
\begin{array}{c}
 a_n^{(2)}(2\om)\\
 b_n^{(2)}(2\om)
\end{array}
\right)
\int\limits_V  {\vec{W}}^{(1)}_{^e_o mn}(k_2(2\omega),\vec r) \te\chi \vec E^{\omega}(\vec r) \vec E^{\omega}(\vec r) dV\:.
$$
Here, $ {\vec{W}}^{(1)}_{^e_o mn}(k_2(2\omega),\vec r)$ is a  vector spherical harmonic ${\vec{M}}^{(1)}_{^e_o mn}(k_2(2\omega),\vec r)$ or ${\vec{N}}^{(1)}_{^e_o mn}(k_2(2\omega),\vec r)$.

\newpage

\section{Rigorous approach for the selection rules imposed by the spherical symmetry}
\label{scalar_prod_app}
 Considering three scalar products of vector spherical harmonics in Eq.~(\ref{eq:I2}), we find  that each of them can be expanded in a  finite sum of several specific scalar functions $\psi_{^e_omn}$ with the coefficients independent on angle \cite{DeBeer2009, VMK}.  In turn, the product of three scalar functions can be again represented in the form of a sum over spherical scalar harmonics.  We are looking for the cases when the coefficient before the invariant term $\psi_{e00}$ is non-zero, which also provides integral \eqref{eq:I2} is non-zero.  

 Here, we are only interested in the scalar products with Cartesian  harmonics $\vec N_{\alpha}$, which have  angular momentum $n=1$. As a result, we have two different types of scalar products:
\begin{enumerate}
\item If  $\bm W_{\mu n }$ is replaced by $\vec N_{p_rmn}$, and   $\vec N_{\alpha}$ is replaced by $ \vec N_{p_r'm'1}$, where $p_r$ is the parity index, introduced in \ref{parity_paragraph}, obtaining \cite{Stout2006}:
       \begin{gather}
 \left[ \vec N_{p'm'1} \vec N_{pmn}\right] \propto\sum_{^{m''=m\pm m',}_{^{ n''=n\pm 1,}_{  n''\neq n}}} c(r) \psi_{p\cdot p'm''n''}.
  \end{gather} 
 For instance:
  \begin{gather}
[{\vec{N}}_{e01}(0, \vec r)\cdot{\vec{N}}_{e01}(k, \vec r))] =[\vec{N}_{e01}]_z  = \\
=\frac23\left( \frac{z_1(\rho)}{\rho} \left(\frac{\psi_{e00}}{z_0(\rho)}+\frac{\psi_{e02}}{z_2(\rho)}\right)+\frac{z_0(\rho)-z_2(\rho)}{3}\left(\frac{\psi_{e00}}{z_0(\rho)}-\frac{\psi_{e02}}{z_2(\rho)}\right)\right).
\end{gather}

\item If $\bm W_{\mu n }$ is replaced by  $\vec M_{p_rmn}$, and   $\vec N_{\alpha}$ is replaced by $ \vec N_{p_r'm'1}$, we obtain a similar expression. Since  for Cartesian projections of magnetic harmonics $n''=n$, the  the summation over full angular momentum is simplified:
          \begin{gather}
 \left[ \vec N_{p'm'1} \vec M_{pmn}\right] \propto\sum_{^{m''=m\pm m', }_{n''=n}} c(r) \psi_{p\cdot p'm''n''}.
  \end{gather}

  For instance: 
\begin{gather}
[{\vec{N}}_{o11}(0, \vec r)\cdot{\vec{M}}_{o11}(k, \vec r)] = [\vec{M}_{o11}]_y  \sim \psi_{o01} = 0
\end{gather}
\begin{gather}
[{\vec{N}}_{e01}(0, \vec r)\cdot{\vec{M}}_{e13}(k, \vec r)] =[\vec{M}_{e13}]_z  = \frac23 z_3(\rho)\psi_{o13}
\end{gather}
\end{enumerate}
\blue{In our considerations, we are not interested in the exact form of the coefficients $c(r)$, because they have no angular dependence, so they are invariant under all transformations of the sphere and can't alter the selection rules. If radial integration turns into zero, this is not due to the symmetry and can't be considered in this simple way. } Finally, we obtain  the integrand  consisting of the sum of products of $scalar$ spherical harmonics $\psi_{^e_omn}$, which  can be easily expressed via the Clebsh-Gordan coefficients \cite{Dong2002, James1976}. This means that the product of two of three  scalar harmonics must contain a third one:
 
       \begin{gather}
\psi_{p'm'n'}\psi_{pmn}\propto\sum_{m''=m\pm m', n''} \psi_{p\cdot p'm''n''} C^{n''0}_{n0n'0} \label{scalarsum}.
  \end{gather} 

The Clebsh-Gordan coefficient $C^{n''0}_{n0n'0}$ is non-zero only when $n''$ has the same parity as sum of $n$ and $n'$. The usual triangle inequality for $n$ must be satisfied as well. It appears in accordance with the fact that product of two functions should have the same inversion behavior as third.
 
Below we present several   examples of computation of the integrals, obtaining selection rules for particular mode channels:
 
        \begin{gather}
\int\limits_V \chi_{xxz}\left[ \vec N_x \vec M_{o11}(2\om)\right] [\vec N_x \vec M_{o11}(\om)]\left[\vec N_z \vec N_{e01}(\om)\right]dV =\\
=\int\limits_V \chi_{xxz}\left[ \vec N_{e11}(0) \vec M_{o11}(2\om)\right] [\vec N_{e11}(0) \vec M_{o11}(\om)]\left[\vec N_{e01}(0) \vec N_{e01}(\om)\right]dV \rightarrow \\
\rightarrow \chi_{xxz} \int\limits_V \psi_{e01} \psi_{e01} (c_1\psi_{e02}+c_2\psi_{e00}) dV \neq 0,
 \end{gather} here $c_1$ and $c_2$ depend on the radius only and are angular-independent.
Thus, the coupling is possible with the tensor component $\chi_{xxz}$.
\\

           \begin{gather}
\int\limits_V \chi_{xxz}\left[ \vec N_x(0) \vec N_{e11}(2\om)\right] [\vec N_x(0) \vec M_{o11}(\om)]\left[\vec N_z(0) \vec N_{e01}(\om)\right]dV =\\
=\int\limits_V \chi_{xxz}\left[ \vec N_{e11}(0) \vec N_{e11}(2\om)\right] [\vec N_{e11}(0) \vec M_{o11}(\om)]\left[\vec N_{e01}(0) \vec N_{e01}(\om)\right]dV \rightarrow \\
\rightarrow \chi_{xxz} \int\limits_V (c_1\psi_{e22} +c_2 \psi_{e00}+c_3\psi_{e02})\psi_{e01} \psi_{e02} dV = 0,
 \end{gather}
 so this coupling is prohibited because $C^{20}_{1020}$ is zero. We see that actually, it is prohibited due to   Rule A because of the integrand is odd with respect to spatial inversion. The three selection Rules A-C, given in the main text, follow from this procedure, but mix all the harmonics, neglecting properties of specific scalar products.
 
 \section{Symmetry classification of vector spherical harmonics.}
 \label{app:hrmncs_symm_table}
In order to reveal how the crystalline symmetry affects the possibility of multipolar generation, we need to know the behavior of  the vector spherical functions under the transformations from the crystal symmetry group. Here we give the table for two types of crystalline symmetries where we indicate  the corresponding irreducible representations  and express the spherical functions via the basis functions of these representations, that are transformed via each other in the same way. The numerical coefficients are obtained properly, but their explicit values are not required  to derive the selection rules. While the   selection rules in spherical  BaTiO$_3$ nanoparticles  can be determined  just from the conservation of the angular momentum projection quantum number $m$,  the table can  be also useful for the nanoparticles of the pyramidal shape. \blue{Similar classifications for other symmetries can be found in \cite{PhysRevB.98.165110}}
\begin{table}[ht]
\caption{Symmetry classification of vector spherical harmonics}  \label{table:symm_irr_reps}
\begin{center}
\begin{tabular}{|cc|lc|lc|}
\hline\multicolumn{2}{|c|}{Spherical Harmonic}& \multicolumn{2}{c}{AlGaAs ($T_{d}$)}\vline & \multicolumn{2}{c}{BaTiO$_{3}$ ($C_{4v}$)}\vline\\\hline
\multirow{3}{*}{ED}&$\bm N_{o11}$&\multirow{3}{*}{$F_{2}$}&$y$&\multirow{2}{*}{$E$}&$y$\\
&$\bm N_{e11}$&&$x$&&$x$\\\cline{5-6}
&$\bm N_{e01}$&&$z$&$A_{1}$&$z$\\\hline
\multirow{3}{*}{MD}&$\bm M_{o11}$&\multirow{3}{*}{$F_{1}$}&$L_{y}$&\multirow{2}{*}{$E$}& {$x \ \text{or} \  L_y$}\\
&$\bm M_{e11}$&&$L_{x}$&&{$y \ \text{or} \ L_x$}\\\cline{5-6} 
&$\bm M_{e01}$&&$L_{z}$&$A_{2}$&$L_{z}$\\\hline
\multirow{5}{*}{EQ}&$\bm N_{o12}$&\multirow{3}{*}{$F_{2}$}&$x$&\multirow{2}{*}{$E$}&$\blue{y}$\\
&$\bm N_{e12}$&&$y$&&$\blue{x}$\\\cline{5-6}
&$\bm N_{o22}$&&$2z$&$B_{2}$&$xy$\\\cline{3-6}
&$\bm N_{e22}$&\multirow{2}{*}{$E$}&$6(x^{2}-y^{2})$&$B_{1}$&$x^{2}-y^{2}$\\\cline{5-6}
&$\bm N_{e02}$&&${2z^2-x^2-y^2}$&$A_{1}$&$z$\\\hline
\multirow{5}{*}{MQ}&$\bm M_{o12}$&\multirow{3}{*}{$F_{1}$}&$L_{x}$&\multirow{2}{*}{$E$}&\blue {$x \ \text{or} \  L_y$}\\
&$\bm M_{e12}$&&$L_{y}$&& \blue{$y \ \text{or} \  L_x$}\\\cline{5-6}
&$\bm M_{o22}$&&$L_{z}$&$B_{1}$&$x^{2}-y^{2}$\\\cline{3-6}
&$\bm M_{e22}$&\multirow{2}{*}{$E$}&${2z^2-x^2-y^2}$&$B_{2}$&$xy$\\\cline{5-6}
&$\bm M_{e02}$&&$6(x^{2}-y^{2})$&$A_{2}$&$L_{z}$\\\hline
\multirow{7}{*}&$\bm N_{o13}$&\multirow{2}{*}{$F_{2}+F_1$}&${-6y+L_y}$&\multirow{2}{*}{$E$}&$y$\\
&$\bm N_{e13}$&&${-6x-L_x}$&&${x}$\\\cline{3-6}
&$\bm N_{o23}$&$A_1$&$1$ or \ $ xyz$  &$B_{2}$&$xy$\\\cline{3-6}
{EO}&$\bm N_{e23}$&$F_1$&$4 L_z$&$B_{1}$&$x^{2}-y^{2}$\\\cline{3-6}
&$\bm N_{o33}$&\multirow{2}{*}{${F_1+F_2}$}&${-60y-6L_y}$&\multirow{2}{*}{$E$}&$y$\\
&$\bm N_{e33}$&&${60{x}-6L_x}$&$$&$x$\\\cline{3-6}
&$\bm N_{e03}$&$F_2$&$4z$&$A_{1}$&$z$\\\hline
\multirow{7}{*}&$\bm M_{o13}$&\multirow{2}{*}{$F_{2}+F_1$}&${-6L_y+y}$&\multirow{2}{*}{$E$}&$L_y$\\
&$\bm M_{e13}$&&${-6L_x-x}$&&${L_x}$\\\cline{3-6}
&$\bm M_{o23}$&$A_2$&${1*}$ or $L_xL_yL_z$&$B_{1}$&$x^{2}-y^{2}$\\\cline{3-6}
{MO}&$\bm M_{e23}$&$F_2$&$4z$&$B_{2}$&$xy$\\\cline{3-6}
&$\bm M_{o33}$&\multirow{2}{*}{${F_1+F_2}$}&${-60L_y-6y}$&\multirow{2}{*}{$E$}&$L_y$\\
&$\bm M_{e33}$&&${60{L_x}-6x}$&$$&$L_x$\\\cline{3-6}
&$\bm M_{e03}$&$F_1$&$4L_z$&$A_{2}$&$L_z$\\\hline
\end{tabular}
\end{center}
\end{table}

\bluee{\section{Anisotropy of the linear material parameters}}

\begin{figure}[htbp]
\includegraphics[width=0.5\linewidth]{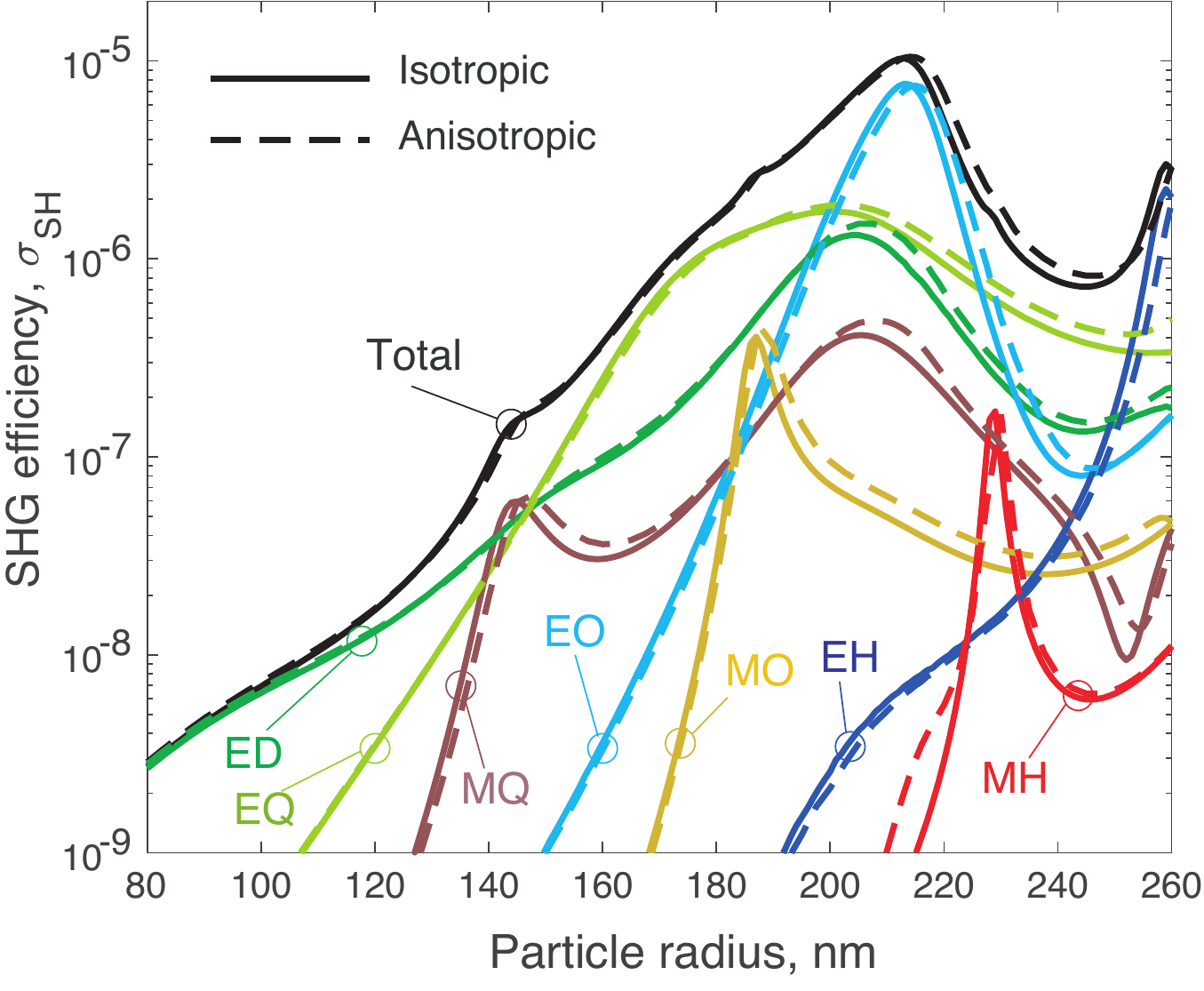}
\caption{\bluee{The SHG cross section for BaTiO$_3$ nanoparticle with account for the anisotropy of material parameters. All the parameters of simulations are the same as in Fig.~\ref{btmult} (a), but $\sqrt{\eps_{xx}}=\sqrt{\eps_{yy}}=2.48$ and $\sqrt{\eps_{zz}}=2.42$ at the SH  wavelength of  525~nm.}}
\label{fig:aniso}
\end{figure}

 \end{widetext}

\end{document}